\documentclass[twocolumn,amsmath,amssymb]{revtex4-1}

\usepackage{mathrsfs}
\usepackage{mathtools}
\usepackage{xfrac}
\usepackage{color}

\usepackage{graphicx}
\usepackage{dcolumn}
\usepackage{bm}

\newcommand{\anni}[2]{\hat{\psi}_{#1}(#2)}
\newcommand{\crea}[2]{\hat{\psi}^{\dagger}_{#1}(#2)}

\newcommand{\low}{\hat{a}}
\newcommand{\rais}{\hat{a}^{\dagger}}

\newcommand{\ket}[1]{ | #1 \rangle }

\newcommand{\order}[1]{\langle #1 \rangle }

\newcommand{\vet}[1]{\vec{#1}}

\newcommand{\up}{\uparrow}
\newcommand{\down}{\downarrow}

\usepackage{hyperref}

\begin{document}


\title{Interacting Dirac fermions under spatially alternating pseudo-magnetic field:  \\ Realization of spontaneous quantum Hall effect}

\author{J\"orn W. F. Venderbos}
\affiliation{%
Department of Physics, Massachusetts Institute of Technology, Cambridge, Massachusetts 02139, USA
}%

\author{Liang Fu}
\affiliation{%
Department of Physics, Massachusetts Institute of Technology, Cambridge, Massachusetts 02139, USA
}%

\date{\today}

\begin{abstract}
Both topological crystalline insulators surfaces and graphene host
multi-valley massless Dirac fermions which are not pinned to a
high-symmetry point of the Brillouin zone. Strain couples to the
low-energy electrons as a time-reversal invariant gauge field, leading
to the formation of pseudo-Landau levels (PLL). Here we study periodic
pseudo-magnetic fields originating from strain superlattices. 
We study the low-energy Dirac PLL spectrum induced by the strain
superlattice and analyze the effect of various polarized
states. Through self-consistent Hartree-Fock calculations we establish
that, due to the strain superlattice and PLL electronic structure, a
valley-ordered state
spontaneously breaking time-reversal and realizing a quantum Hall phase is favored, while others are suppressed. Our analysis applies to both topological
crystalline insulators and graphene. 
\end{abstract}

\maketitle


\section{Introduction\label{sec:intro}}

The discovery of graphene and topological insulators has significantly boosted the ubiquity of condensed matter
realizations of Dirac fermions as emergent electronic excitations at low-energy~\cite{novoselov04, hasan10,qi11}. 
Dirac electrons in condensed matter systems have
enjoyed an enormous amount of interest both from a fundamental
and technological application perspective~\cite{castro09}. 
A key difference between graphene and topological insulators is the
number of species, or valleys, of Dirac fermions and their locations in momentum space. 
Topological insulators (TI) protected by time reversal symmetry host a single-valley Dirac fermion, which is pinned to a time-reversal-invariant (TRI) 
momentum  in the surface Brillouin zone~\cite{fu07}. 
In contrast, graphene hosts two valleys of Dirac fermions located at non-TRI momenta~\cite{semenoff84,haldane88}, 
each valley having an additional spin degeneracy. More recently, a new type of Dirac fermions was discovered on the surface of 
topological crystalline insulators (TCI) SnTe, (Sn,Pb)Se and (Sn,Pb)Te~\cite{hsieh12, ando, poland, hasan, AndoFu}, which are protected by mirror symmetry of the crystal~\cite{hsieh12, vidya1,vidya2,serbyn14}. 
These Dirac fermions exhibit spin-momentum locking as in TIs; however, there is an even number of Dirac cones at non-TRI momenta, a feature similar to graphene. 

In general, when Dirac points are located at non-TRI momenta, nonmagnetic perturbations such as strain 
are able to move Dirac points in momentum space, thereby acting as an effective gauge field on Dirac fermions. 
For example, strain induces opposite gauge fields for the two Dirac valleys in graphene, and spatially
inhomogeneous strain gives rise to effective magnetic fields that are opposite in two valleys, preserving time reversal symmetry~\cite{guinea10,vozmediano10,amorim15}. In the presence of such a pseudo-magnetic
field $\mathcal{B}$, the low-energy electronic structure takes the form of
pseudo-Landau levels (PLLs) with energies characteric of Dirac electrons in
magnetic fields, i.e. $\sim \sqrt{n\mathcal{B}}$, where
$n$ is the Landau level (LL) index. Key signatures of pseudo-magnetic
fields have been experimentally observed in
graphene~\cite{gomes12,levy10}.

The PLLs have a large single-particle degeneracy,
which makes them susceptible to many-body instabilities in a manner
similar to magnetic-field induced LLs~\cite{gusynin94}. Electronic
interactions are expected to lift the degeneracy and drive the system into various gapped states. 
Two primary examples are spin-polarized and
valley-polarized states of PLLs in graphene~\cite{herbut08,ghaemi12,abanin12,roy13,roy14a,roy14b}. 


In this work we consider interacting Dirac electrons under \emph{periodically} modulated pseudo-magnetic fields, where  
regions of positive and negative fields alternate in space, forming a superlattice.  
This field profile leads to a novel electronic structure
markedly different from uniform pseudo-magnetic
fields~\cite{vozmediano08,tang14}. There are various ways in which
periodic pseudo-magnetic fields can arise, one prominent way being a strain
superlattice in graphene or TCI. Such
spatially periodic strain fields are particularly relevant, as they
were experimentally found to develop at interfaces of
heterostructures built from TCIs (e.g., SnTe) and trivial insulators (e.g., PbTe)~\cite{sipatov09,palatnik81,springholz01}. At these
interfaces, the lattice constant mismatch causes dislocations which self-organize into a periodic array and therefore produce a natural realization of
periodic strain fields. The key characteristic of the corresponding periodic
pseudo-magnetic fields is that they can exist over macroscopic
regions. In contrast, uniform pseudo-magnetic fields cannot exist in the thermodynamic limit, owing to the boundedness of strain (=pseudo-gauge field)~\cite{tang14}, unlike a real magnetic field. 
Apart from
strain superlattices, periodic pseudo-magnetic fields can arise as a
result of incommensurate electrostatic potentials
originating from, for instance, from lattice-mismatched substrates
with a twist angle~\cite{gopa12,juan13,wallbank13}. 

Starting from the strain-induced pseudo-magnetic superlattice, we address the effect of
electron-electron interactions. 
Spatially alternating pseudo-magnetic fields change the low-energy electronic structure
close to the Dirac points. Most strikingly, the energy-momentum dispersion in the vicinity of {\it each} Dirac point becomes nearly flat, leading to a segment of flat band with a {\it twofold} degeneracy.   
These flat bands arise from the zeroth PLL in regions with strong pseudo-magnetic fields; and the twofold degeneracy corresponds to  
Landau orbitals that reside in different spatial regions of opposite fields and have opposite Dirac spinor components~\cite{tang14}, 
a novel feature that is absent in the case of uniform magnetic fields. Counting the two valleys, the flat bands have fourfold degeneracy.  
The presence of flat bands leads to a diverging density of states, in contrast to the vanishing density of states at the
Dirac point of massless Dirac fermions. Consequently, periodic strain fields provide a feasible and effective
way of engineering density of states, i.e., electronic compressibility, at zero energy.

At charge neutrality, the degenerate flat bands are mainly responsible for driving the
spontaneous formation of ordered states. We discuss the various
possibilities for degeneracy lifting in the flat band and discriminate
between energetically favorable and unfavorable states. Two prominent
candidate ordered states are the charge-ordered state, where charge
is redistributed from the region of positive (negative) to negative
(positive) pseudo-magnetic field, and the valley-ordered state,
where in each spatial region the valley degeneracy is lifted. 
We show that the valley-ordered state in graphene and TCIs spontaneously breaks time-reversal symmetry and realizes  
an integer quantum Hall effect similar to the Haldane state~\cite{haldane88}, 
but with the important difference of being driven by electron interactions without any external time-reversal-breaking field.  

We determine the mean-field ground state by self-consistently solving
the full gap equation of interacting Dirac fermions under a periodically alternating pseudo-magnetic field. The continuum Hamiltonian is microscopically implemented 
using a lattice model with a strain superlattice.  
Our analysis shows that the support of the flat band wavefunctions is of great importance. Flat bands in any spatial region only have a twofold valley degeneracy, protected by 
the time reversal symmetry. Therefore lifting this degeneracy by interactions implies time reversal symmetry breaking. For this reason, 
we find the valley-ordered quantum Hall state is  greatly favored over the charged-ordered state under generic forms of electron interactions. We show that the order
parameters corresponding to the ordered states follow the strain
profile, highlighting the crucial role of the pseudo-magnetic field.

The present setup for a spontaneous time-reversal symmetry breaking quantum Hall state 
relying on a strain-induced flat band should be
contrasted with the proposed Haldane mass generation for interacting massless Dirac fermions
in graphene~\cite{raghu08,weeks10,grushin13}. Exact
diagonalization and DMRG
studies~\cite{daghofer14,garcia13,duric14,motruk15} seem to have failed to find the interaction-driven Haldane phase in models so far proposed, 
in contradiction to Hartree-Fock results. It is believed that the absence of the Haldane phase in ED and DMRG
phase diagrams stems from the vanishing density of states at the Dirac point, and the resulting absence of a weak-coupling instability. 
In contrast, the quantum Hall state in our setup already occurs spontaneously at weak coupling, owing to the strain-induced flat band.






\section{Dirac fermions in two dimensions\label{sec:2ddirac}}

We set out to study the coupling of time-reversal invariant
pseudo-gauge fields to Dirac electrons. With two specific realizations
in mind, graphene and TCI surface states, we focus on Dirac electrons
in two dimensions. Let us start by stating the essential features of
Dirac electrons coupled to pseudo-gauge fields, independent of
specific context.

Any two dimensional system respecting
time-reversal invariance and having Dirac fermions not pinned to a
particular time-reversal invariant momentum, will consist of two
species of Dirac fermions. Labeling the two species by $+$ and $-$,
the Dirac Hamiltonian describing the two species takes the general form
\begin{gather} \label{eq:diracham}
\hat{\mathcal{H}}_\pm = \pm \hbar v_F \hat{\Psi}^\dagger_{\pm} (-i
\tau^x\partial_x - i\tau^y\partial_y  ) \hat{\Psi}_{\pm}
\end{gather}
where $\tau^i$ is a set of Pauli matrices acting on the pseudospin
degree of freedom of the Dirac fermions. Time-reversal symmetry
relates the two species by exchanging $ \hat{\Psi}_{+} \leftrightarrow  \hat{\Psi}_{-} $.  

Pseudo-gauge fields couple to the Dirac fermions in a manner similar
to real electromagnetic gauge fields, with one crucial
difference, however. In order to respect time-reversal invariance, the pseudo-gauge field
must couple to the fermions in such a way that the two species see opposite
fields. As a result, in the presence of a pseudo-gauge field given by
$\mathcal{A}_\mu$ ($\mu=x,y$), the Dirac Hamiltonian becomes
\begin{gather} \label{eq:dirachamfield}
\hat{\mathcal{H}}_\pm = \pm  v_F \hat{\Psi}^\dagger_{\pm}
\tau^\mu (\hat{p}_\mu \pm \mathcal{A}_\mu ) \hat{\Psi}_{\pm}.
\end{gather}
This Hamiltonian (with $\hat{p}_\mu = -i \partial_\mu$) describes the generic Dirac electrons
coupled to pseudo-gauge fields. Pseudo-Landau level quantization will
occur when the gauge field $\mathcal{A}_\mu$ acquires spatial
dependence, i.e., $\mathcal{A}_\mu = \mathcal{A}_\mu(\vec{r})$.

The interpretation of the Dirac
fermion pseudospin and valley
degrees of freedom will depend on the particular realization of
pseudo-magnetic field coupling in a given material. In this work we will discuss two examples of low-energy Dirac electrons
coupled to time-reversal invariant gauge fields, which we introduce in
the remainder of this section. First we consider the case of graphene, and
then we consider surface states of TCIs. Whereas in graphene the Dirac
pseudospin degree of freedom derives from the two sublattices~\cite{semenoff84}, the
pseudospin of the TCI surface state is more complicated due to intrinsic
spin-orbit coupling, as we will discuss below.
Importantly, in case of the latter, spin-orbit coupling leads to
spin-momentum locking in the surface state Dirac theory. 

In both cases, graphene and TCI, the emphasis will be on strain-induced pseudo-magnetic
field coupling. However, we will use the case of graphene to point out
that pseudo-magnetic fields can have a physical origin different from
strain, giving way to an even wider application of our results.

\subsection{Dirac fermions in graphene\label{ssec:tcisurface}}

The low-energy theory of graphene at charge neutrality is one
of the hallmark
examples of a $2D$ Dirac
theory~\cite{semenoff84,beenakker08,castro09}. The
two species of nodal Dirac fermions are located at the two inequivalent BZ
corners, i.e., the Dirac points or
valleys, and are labeled by $K_+$ and $K_-$ corresponding to the momenta
$\vec{K}_+   = (4\pi/3,0)$ and $\vec{K}_-   = -(4\pi/3,0)$. The Dirac
Hamiltonian is obtained by expanding
the band structure around the Dirac points $K$ and $K'$ in small
momenta $\vec{q}$ relative to the Dirac
points~\cite{semenoff84}. It is given by
\begin{gather} \label{eq:graphene}
\mathcal{H}(\vec{q}) = \hbar v_F \nu^z ( q_x\tau^x + q_y \tau^y) \equiv \hbar
v_F q_\mu \Gamma_\mu 
 \end{gather}
(where $v_F = \sqrt{3}ta/(2\hbar)$). 
The set of Pauli matrices $\tau^i$ acts on the
sublattice degree of freedom ($A$/$B$) and the set of matrices $\nu^i$ acts
on the valley degree of freedom ($K_+$/$K_-$ ). In addition, we have defined the Dirac matrices $\Gamma_x = \nu^z\tau^x$ and
$\Gamma_y = \nu^z\tau^y$. 
The Hamiltonian acts on the Dirac spinor $\hat{\Psi}(\vec{q})$ defined by
\begin{gather} \label{eq:diracspinor}
\hat{\Psi}(\vec{q}) = \begin{bmatrix} \anni{A +}{\vec{q}}  & \anni{B +}{ \vec{q}} &  \anni{B -}{\vec{q}}  & \anni{A - }{ \vec{q}}  \end{bmatrix}^T.
\end{gather}
Note that we choose the basis so that the $A$ and $B$ lattice are
exchanged in the $K_-$ valley, meaning
that we are working in the chiral representation (i.e., $
\mathcal{H}_{\pm} = \pm \vec{q}\cdot \vec{\tau}$ for valley $K_\pm$). 

Starting from the low-energy Dirac Hamiltonian of
Eq.~\eqref{eq:graphene}, we introduce a generalized
time-reversal invariant pseudo-gauge field by
coupling the Dirac fermions to the field
\begin{gather} \label{eq:gf}
\vec{\mathcal{A}}^i = (\mathcal{A}^i_x,\mathcal{A}^i_y),
\end{gather}
which consists of three components $i=1,2,3$. The coupling to the
fermions has the same form as 
ordinary minimal coupling, but with different gauge charges
$\Omega^i$ expressed as
\begin{gather} \label{eq:gfcoupling}
\mathcal{H}(\vec{q}) = \hbar v_F \Gamma_\mu (q_\mu +
\mathcal{A}^i_\mu \Omega^i).
\end{gather}
The gauge charge matrices $\Omega^i$ encode the distinct nature of the
pseudo-gauge field as compared to the ordinary gauge field, and are given by
$\Omega^i = (\nu^x\tau^z,\nu^y\tau^z,\nu^z  )$. The third gauge charge
matrix $\Omega^3 = \nu^z $ is diagonal in valley space and assigns
opposite sign to the two valleys. Therefore, the component $\mathcal{A}^3_\mu
\Omega^3$ realizes the general Hamiltonian of
Eq.~\eqref{eq:dirachamfield} in graphene. In graphene, this is the pseudo-gauge
field component that arises in the presence of strain and plays a
central role in this work. The presence of a field $\vec{\mathcal{A}}$
coupling to $\Omega^3$ leads to a moving of the Dirac points away from
$K_+$ and $K_-$, in opposite directions. This is shown in
Fig.~\ref{fig:bz} (left), where the bold blue dots denote the Dirac
points moving towards the zone center. 

The following properties of the gauge field charges will be important for our analysis. The charges $\Omega^i$ realize a pseudospin
$SU(2)$ algebra, expressed as $[\Omega^i, \Omega^j  ] = 2 i \epsilon^{ijk}
\Omega^k$. The matrices $\Omega^i$ commute with the
Hamiltonian in the absence of fields, and as a consequence generate a continuous
$SU(2)$ symmetry of the low-energy graphene Hamiltonian. This symmetry
is broken when mass terms are introduced to the Hamiltonian, i.e., when
the Dirac electrons are gapped out. In particular, the set of mass
matrices $\vec{\Gamma}
= (\Gamma_1,\Gamma_2,\Gamma_3) \equiv (\nu^x,\nu^y,\nu^z\tau^z)$
describes masses that anti-commute with the Hamiltonian and between
themselves. They constitute a set of compatible masses, the
physical nature of which is well-known. Specifically, the
mass $\Gamma_3 = \nu^z\tau^z$ corresponds to an electrostatic potential making the
two honeycomb sublattices inequivalent and breaking inversion
symmetry. Such term is diagonal in valley-space, i.e., it does not
couple the two Dirac points. The other two masses, $\Gamma_1$ and
$\Gamma_2$, which are off-diagonal in valley space, are known as Kekul\'e masses and correspond to modulations
of the tight-binding nearest-neighbor hopping parameter $t$ with
tripled unit cell~\cite{hou07}. The breaking of translational
invariance and the modulations over small distances (large
momenta) couple the Dirac points. 

The gauge
charges $\Omega^i$ act as generators of rotations within the space of
masses, which follows from the commutation relation $[\Omega^i, \Gamma_j  ] = 2 i \epsilon^{ijk}
\Gamma_k$.  In addition to the mass terms $\vec{\Gamma}$,
there is a mass term $\tau^z$, the time-reversal odd
Haldane mass~\cite{haldane88}, which anti-commutes with
the Hamiltonian~\eqref{eq:graphene}, but commutes with both the 
$\Gamma_i$ and the $\Omega^i$. Hence, whereas $\vec{\Gamma}$ is a
vector under the transformations generated by $\Omega^i$, $\tau^z$ is
a scalar. 

The two remaining gauge charges $\Omega^1 = \nu^x\tau^z $ and $\Omega^2 = \nu^x\tau^z$ are off-diagonal in
valley space but diagonal in sublattice space. The former implies
translational symmetry breaking and the latter implies that these
terms arise due to charge density modulations. Consequently, charge density
waves (CDWs) with a six-site unit cell, which we will refer to as valley-coupling CDWs, lead to a pseudo-gauge coupling in the same
way as strain~\cite{gopa12}. The $SU(2)$ structure of
the gauge charges $\Omega^i$ implies that within the low-energy
theory, the pseudo-gauge field components are unitarily equivalent to
each other.

\subsection{TCI surface state Dirac fermions\label{ssec:tcisurface}}

Topological insulator materials are bulk insulators hosting gapless
Dirac fermions at their surfaces~\cite{hasan10,qi11}. The
spin-momentum locked surface Dirac fermions are protected by
time-reversal symmetry, and as a result they are pinned at the time-reversal invariant
momenta (TRIM). Due to this symmetry-protected pinning, the surface
states of topological insulators do not allow for time-reversal invariant pseudo-gauge field
coupling. In particular, strain is not able to move the Kramer's doublet away
from the TRIM. 

\begin{figure}
\includegraphics[width=\columnwidth]{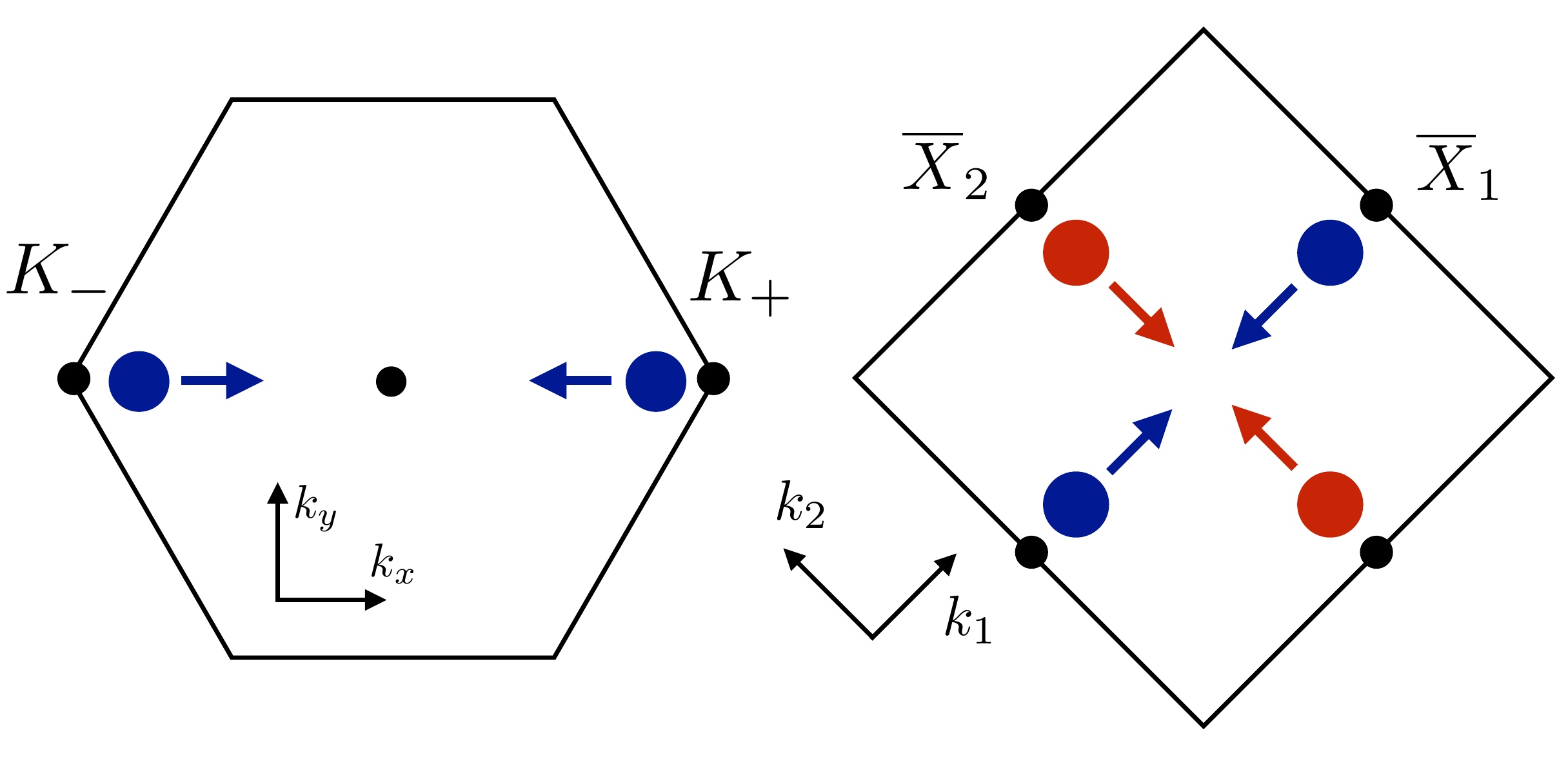}
\caption{\label{fig:bz}  (Left) Hexagonal Brillouin zone of the
  honeycomb lattice. The two Dirac points $K$ and $K$ are marked by
  bold blue dots. The blue arrows indicate possible Dirac points moving towards
the zone center due to strain, i.e. $\sim u_{xx}-u_{yy}$. (Right) Square
surface Brillouin of TCI surface state with two sets of Dirac points
located at $\overline{X}_1$ (blue) and $\overline{X}_2$ (red). Arrows
indicate the moving of Dirac points towards the zone center due to
symmetric strain $\sim  u_{xx}+u_{yy}$.}
\end{figure}

In contrast, the TCIs are topological materials protected by crystalline symmetries~\cite{fu11,hsieh12}, which host
surface Dirac fermions not pinned to the
TRIM~\cite{liu13,serbyn14,fang15}. As a result, strain
can couple to the low-energy Dirac fermions as a pseudo-gauge field
and can move the Dirac points in momentum space, in a way that depends on the symmetry
of the strain tensor~\cite{serbyn14,tang14}. 

In this work we
specfically focus on the SnTe material class~\cite{hsieh12} and
its mirror symmetry-protected surface Dirac
fermions appearing on the $(001)$ surface. The surface Brillouin zone
of the $(001)$ surface is shown in Fig.~\ref{fig:bz}. Two species of
low-energy Dirac fermions related by time-reversal symmetry exist in the
vicinity of the surface time-reversal invariant momenta
$\overline{X}_1$ and $\overline{X}_2$, represented as blue and red
dots in Fig.~\ref{fig:bz}. The surface state Dirac Hamiltonian at
$\overline{X}_1$, given by the terms that respect the crystal symmetries leaving
$\overline{X}_1$ invariant, reads
\begin{gather} \label{eq:hamtci}
\mathcal{H}_{\overline{X}_1}(\vec{q}) = v_1q_1\sigma^y -
v_2q_2\sigma^x + m \nu^x + \delta \sigma^x\nu^y,
\end{gather}
and a similar expression can be derived for $\overline{X}_2$. Here $\sigma^i$ is a set of Pauli
matrices that represents a Kramers doublet, and $\nu^i$ is a valley degree of freedom
corresponding to the two inequivalent bulk $L$-points mapped onto
$\overline{X}_1$. The momentum $\vec{q}$ is measured with respect to
$\overline{X}_1$; the spin-momentum locking shown in
Hamiltonian~\eqref{eq:hamtci} (i.e., first two terms) comes from 
spin-orbit coupling. For $m=\delta=0$ there are two degenerate Kramer's
doublets at $\overline{X}_1$, which are split in energy by finite $m$
and $\delta$. Most importantly, finite $m$ and $\delta$ leads to the
appearance of two species low-energy Dirac points, which are located at
$\vec{\Lambda}_\pm = (0,\pm\sqrt{m^2+\delta^2}/v_2)$, measured from
$\overline{X}_1$. 

The Hamiltonian of Eq.~\eqref{eq:diractci} can be projected into
the subspace corresponding to $\vec{\Lambda}_\pm$ to obtain the
effective low-energy Dirac theory. This yields~\cite{serbyn14}
\begin{gather} \label{eq:diractci}
\mathcal{H}_{\vec{\Lambda}_\pm}(\vec{q}) = -v'_1q_1\tilde{\tau}^x +
v_2q_2\tilde{\tau}^z,
\end{gather}
where $\tilde{\tau}^i$ is the effective pseudospin degree of freedom,
$\vec{q}$ is now measured with respect to $\vec{\Lambda}_\pm$,
and $v_1' = v_1\delta/\sqrt{m^2+\delta^2}$. Note that in the chosen
basis the Hamiltonian is valley-isotropic, taking $\tilde{\nu}^z=\pm 1$ as
an effective valley degree of freedom. 

With the Hamiltonian of Eq.~\eqref{eq:diractci} we have arrived at a desciption of the low-energy Dirac fermions
that has the general form introduced in the beginning of this section,
and is thus similar to the graphene
Hamiltonian of Eq.~\eqref{eq:graphene}. Hence, in a way analogous to
graphene, we can use symmetry arguments to establish the effect of
various perturbations. For instance, since the TCI surface states are
protected by mirror symmetry, one expects mirror symmetry breaking to
open up a gap. We find two such gap opening mass terms, which do not couple the
low-energy valleys, and they are given in the basis
of~\eqref{eq:diractci} as $\tilde{\tau}^z$ and
$\tilde{\nu}^z\tilde{\tau}^z$. The former is a time-reversal even mass
and corresponds to a ferroelectric distortion of the crystal. It
derives from the Dirac bilinear $\nu^z$ in the basis
of~\eqref{eq:hamtci}. The mass term $\tilde{\nu}^z\tilde{\tau}^z$
breaks time-reversal symmetry and originates from the terms $\sigma^z$
and $\sigma^y\nu^z$ in the basis of Eq.~\eqref{eq:hamtci}. The mass
gap originating from $\tilde{\nu}^z\tilde{\tau}^z$ is equivalent to
the graphene Haldane gap, and consequently corresponds to a QAH phase~\cite{serbyn14}.

Similarly, by using symmetry arguments, the time-reversal invariant pseudo-gauge field
couplings can be identified. As a consequence of the low symmetry of the
$\overline{X}_1$ point, there are no two dimensional representations
which directly imply pseudo-gauge coupling. However, since the symmetric terms
$\nu^x $ and $\sigma^x\nu^y $ displace the Dirac points in momentum
space, any perturbation coupling to them, will have the effect of a pseudo-gauge field.
Looking for other terms both even under time-reversal
and inversion (as expected for strain), one finds another Dirac bilinear given by
$\sigma^y\nu^y$. We will show in the next section, when we discuss strain
and strain superlattices, that components of the
strain tensor couple to these terms. The effect of these terms is shown
schematically in Fig.~\ref{fig:bz} (right), where bold blue and red
dots denote the Dirac points in the vicinity of $\overline{X}_1$ and
$\overline{X}_2$, respectively, shifting towards the zone center as a
result of strain.

\section{Periodic strain superlattices\label{sec:strain}}

In the previous section we introduced pseudo-gauge field coupling in
the context of graphene and TCI surface states, and argued that strain
realizes such coupling. 
We now turn to a more detailed 
discussion of strain, and more specifically \emph{periodic} pseudo-magnetic fields arising due to periodically varying
strain, i.e., a strain superlattice. 

Elastic deformations of the crystal lattice, i.e., strain fields, are described
by the strain tensor $u_{ij}$ given by
\begin{gather}
u_{ij}  = \frac{1}{2}(\partial_i u_j + \partial_j u_i),
\end{gather}
where $u_i$ ($i=x,y$) is the displacement field. Given the symmetry of the crystal lattice, the strain tensor can be decomposed into
components transforming as distinct representations of the symmetry
group. From this decomposition one can read off which lattice deformations couple as
(pseudo-)gauge fields to the Dirac fermions.

\begin{figure*}
\includegraphics[width=0.9\textwidth]{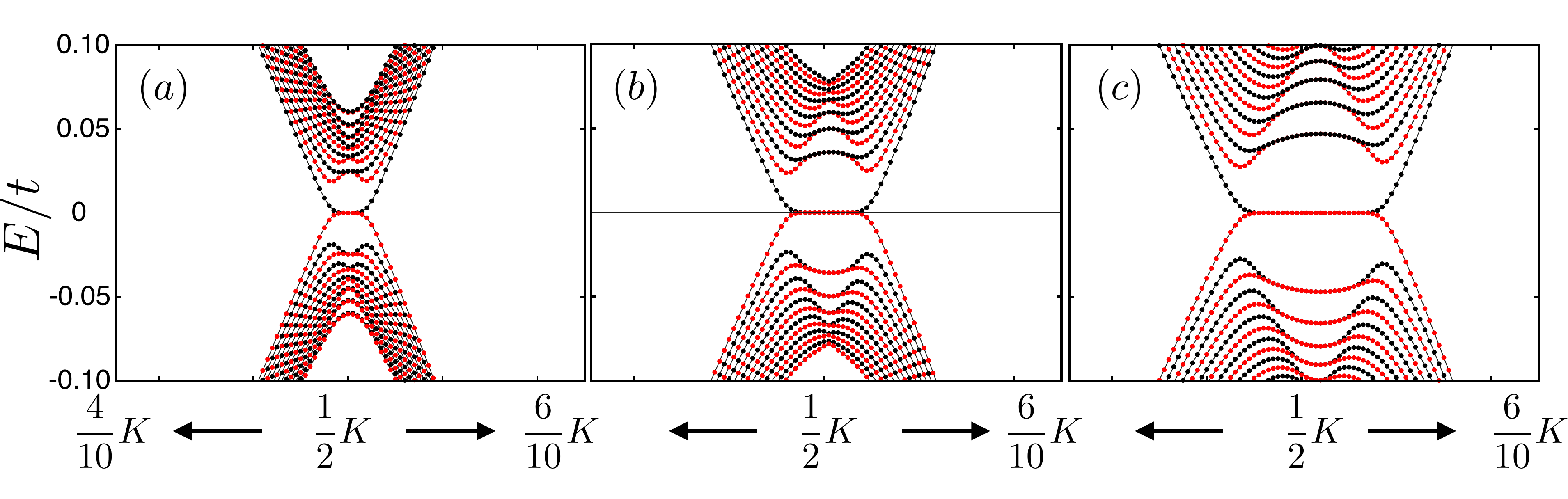}
\caption{\label{fig:spec2} Spectra of graphene in the presence of
  periodically modulated strain for different values of the amplitude
  of modulation (given by $A \sim 2\delta t_1-\delta t_2-\delta
  t_3$; see main text). The modulation wavelength $\lambda$ is chosen to be $700$ graphene unit cells and the
  propagation vector is $\delta \vec{G} = (0, 4\pi/\sqrt{3} ) /
  \lambda$. The amplitudes are (a) $A = 0.01t$, (b) $A = 0.03t$,
  (c) $A = 0.05t$. Note that the plots show $K'$ folded onto the $k_x$-axis.}
\end{figure*}

In case of hexagonal symmetry, applicable to graphene, there are two $d$-wave strain components
$(u_{xx} - u_{yy},  -2 u_{xy}) \sim (d_{x^2-y^2}, d_{xy})$, which
couple to the Dirac fermions as the valley-diagonal field
$\vec{\mathcal{A}}^3$ of Eq.~\eqref{eq:gfcoupling} (let us omit the
label $3$, i.e. $\vec{\mathcal{A}} = \vec{\mathcal{A}}^3$, for the
moment)~\cite{manes07}. Note that assuming full hexagonal symmetry
implies a single coupling constant, 
\begin{gather}
\begin{pmatrix}  \mathcal{A}_x \\ \mathcal{A}_y  \end{pmatrix}
\sim \alpha \begin{pmatrix}  u_{xx} - u_{yy} \\-2 u_{xy}  \end{pmatrix}.
\end{gather}
For illustration purposes, the effect of finite and constant $\mathcal{A}_x$ is shown
graphically in Fig.~\ref{fig:bz} (left), where the Dirac points $K_+$
and $K_-$ move along the $k_x$ axis. In case of the square symmetry, which applies
to the $(001)$ surface states of TCIs, the $d$-wave components
$d_{x_1^2-x_2^2}$ an $d_{x_1x_2}$ are not degenerate, and one finds at $\overline{X}_1$~\cite{tang14},
\begin{gather}
\begin{pmatrix}  \mathcal{A}_x \\ \mathcal{A}_y  \end{pmatrix}
\sim  \begin{pmatrix}  \alpha_1(u_{xx} - u_{yy}) \\ \alpha_2 u_{xy}  \end{pmatrix}.
\end{gather}
In addition, in the previous section we observed that a perturbation
respecting all symmetries can move the Dirac points in momentum space,
implying that $u_{xx} + u_{yy}$ enters the expression for
$\mathcal{A}_x $ as well, with an independent coupling. It is the latter strain component, $u_{xx} + u_{yy}$, the effect
of which is shown in~\ref{fig:bz} (right). 

Let us now come to the case of periodic strain fields with wavelength
$\lambda$. More specifically, we consider $u_{ij}\rightarrow
u_{ij}(\vec{r})$ and consequently $\vec{\mathcal{A}} \rightarrow
\vec{\mathcal{A}}(\vec{r}) $. The periodicity of
$\vec{\mathcal{A}}(\vec{r}) $ is directly reflected in the periodicity
of the pseudo-magnetic field $\mathcal{B} = \vec{\nabla} \times
\vec{\mathcal{A}}(\vec{r})  = \mathcal{B} (\vec{r})$, which should be
compared to and contrasted with uniform
pseudo-magnetic field $\mathcal{B} $. In order to implement the strain superlattice
in a tight-binding setting, we take the graphene lattice as a simple
regularization of the continuum theory. To solve the
superlattice Hamiltonian, we
establish a connection between the strain components and the change in
overlap intergals $\delta t_n$, where $n=1,2,3$ labels the three
nearest neighbor vectors $\{ \vec{\delta}_n \}$. The overlap integral change is expressed in
terms of the strain tensor $u_{ij}$ as $\delta t_n  = \sum_n
\delta^i_n\delta^j_n u_{ij}$, which becomes
\begin{gather}
\begin{pmatrix}  u_{xx} - u_{yy} \\-2 u_{xy}  \end{pmatrix}
\sim \begin{pmatrix}  2\delta t_1 -\delta t_2-\delta t_3\\ \sqrt{3 }(\delta t_2
  - \delta t_3)  \end{pmatrix}.
\end{gather}
This expresses the pseudo-gauge field in terms of the modulation of
the hopping $t_n \rightarrow t+\delta t_n$. 

We proceed to consider a single-propagation
vector pseudo-gauge field modulation and obtain the electronic
spectrum. A particularly convenient choice is the propagation vector $\delta\vec{G} \equiv \vec{G}
/\lambda$, where $\vec{G} = (0, 4\pi/\sqrt{3})$ is a reciprocal
lattice vector. Then $\lambda$ is the superlattice wave length,
given in terms of graphene unit cells, i.e. $\lambda = 700$ leads to a 
superlattice unit cell containing $700$ graphene unit cells. The
pseudo-gauge field $\vec{\mathcal{A}}$ and corresponding
pseudo-magnetic field are given by 
\begin{eqnarray} \label{eq:Mmod}
\mathcal{A}_x(\vec{r}) &=& A \cos (\delta\vec{M}\cdot
\vec{r}) \nonumber \\
\mathcal{B}(\vec{r}) & =& -\partial_y  \mathcal{A}_x  = \delta M_y A \sin (\delta\vec{M}\cdot
\vec{r}),
\end{eqnarray}
while $\mathcal{A}_y = 0$, since we are only interested in the
transverse component. $A$ denotes the amplitude of the
strain field, i.e., the maximal change in overlap $\delta t$. 

The spectra in the presence of the strain superlattice for
some values of $A$ are shown in Fig.~\ref{fig:spec2}. We
observe that for increasing $A$, implying increasing
pseudo-magnetic field strength, a flat zero-energy band forms at the
Dirac points, in addition to higher energy dispersive but doubly
degenerate bands. This specific reorganization of the low-energy
electronic spectrum resembles the Landau level structure of external
magnetic fields. We will establish a detailed connection between Landau
level physics and periodic strain in the next section. A key feature
we wish to stress here, is that the formation of the zero-energy flat band,
the degeneracy of which is related to the strength of the
pseudo-magnetic field, leads to a finite and considerable density of
states (DOS) at the
charge neutrality point. In stark constrast, in the unstrained case Dirac
electrons have linearly vanishing DOS at the charge
neutrality point. 

Instead of the propagation vector $\delta\vec{G}$, we can take the
propagation vector $\delta\vec{K} = \vec{K}/\lambda$, where $\vec{K}$
is the Dirac point vector defined above. This may be viewed as a simple
rotation of $\delta\vec{G}$, which will result in modified Moir\'e pattern. We then have for the
spatially dependent pseudo-gauge field
\begin{eqnarray} \label{eq:Kmod}
\mathcal{A}_x(\vec{r}) &=& A \cos (\delta\vec{K}\cdot
\vec{r}) \nonumber \\
\mathcal{B}(\vec{r}) & =& -\partial_y  \mathcal{A}_x  = \delta K_y A \sin (\delta\vec{K}\cdot
\vec{r})
\end{eqnarray}
where is it important to choose $\vec{K}$ such that the field has a
nonzero transverse component. For given $\lambda$ the strain
superlattice unit cell constains $3\lambda$ graphene unit cells, which
has the benefit that it is commensurate with any perturbation
modulated by $\vec{K}$ coupling the Dirac points. This allows to treat
valley-diagonal and valley-off diagonal perturbations on the same
footing. 

We recall that the low-energy Dirac Hamiltonian in the presence of strain reads $\mathcal{H} = \hbar v_F \Gamma_\mu (-i\partial_\mu +
\mathcal{A}^3_\mu(\vec{r}) \Omega^3)$. As noted above, 
a unitary matrix $U$ can be used to rotate to another gauge field component,
$U^\dagger \mathcal{H} U = \hbar v_F \Gamma_\mu (-i\partial_\mu +
\mathcal{A}^1_\mu(\vec{r}) \Omega^1)$. Clearly, this does not change
the spectrum and therefore electrostatic potential superlattices,
which would couple to the gauge field components $\Omega^1$ and
$\Omega^2$~\cite{wallbank13}, are equivalent to strain superlattices. Therefore, even though we focus on strain in this work, we
highlight that in the
context of graphene spatially modulated valley-coupling CDWs induce
periodic time-reversal invariant pseudo-magnetic fields in the same
way as strain.




\section{Flat band Pseudo-Landau Levels induced by strain\label{sec:PLL}}

\begin{figure}
\includegraphics[width=\columnwidth]{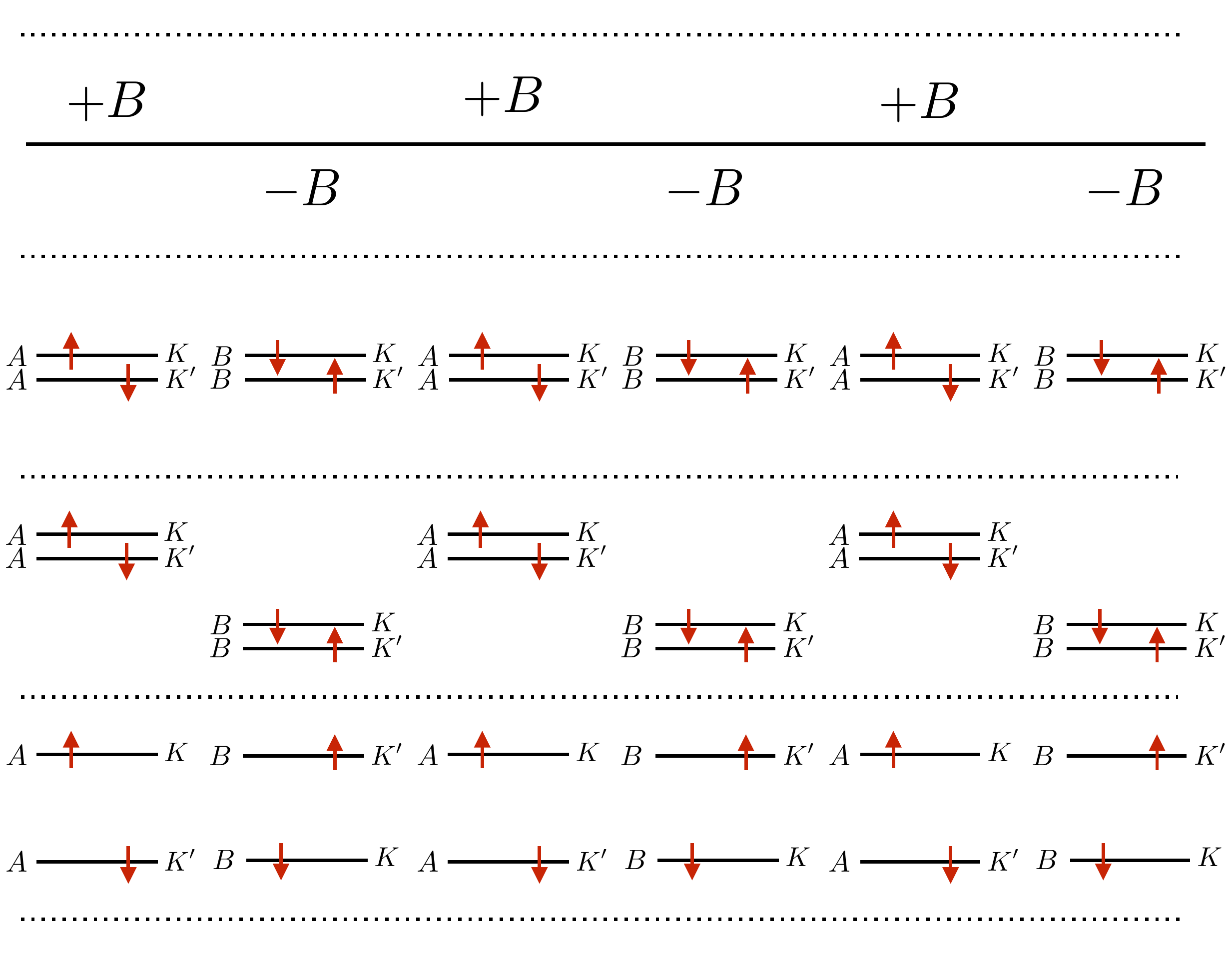}
\caption{\label{fig:pll} Schematic representation of the PLL structure
in case of spatially alternating pseudo-magnetic field. The
alternating regions of positive and negative field are depicted in the
upper row. In the row below we depict the degenerate $n=0$ PLLs in the
two regions, the structure of which periodically alternates in
accordance with the field. Here $A$ and $B$ label a general pseudo-spin
degree of freedom, which corresponds to the $A/B$ sublattice in graphene. The bottom two rows show the two prime
polarized state candidates. In the charge ordered state with
ferroelectric polarization,
the energy levels in the positive field ($A$) have higher
energy than the negative field ($B$) states, the latter
being fully occupied. In the valley-ordered quantum Hall state (very
bottom) levels are split in each region, occupying a single valley in
each region. }
\end{figure}

In this section we address the spectral properties of Dirac electrons
in the presence of a time-reversal invariant pseudo-magnetic field. As in the previous section, we
particularize to the case of graphene (i.e., our lattice regularization of
the continuum theory), and start by considering a spatially uniform
field induced by strain. Uniform fields are fundamentally different
from periodically modulated fields induced by the strain superlattice, but we can use the results for the
former to develop an intuition for the case of periodic pseudo-magnetic
fields. In particular, we may, for the sake of argument, think of the periodic field as alternating regions of
positive and negative constant fields, which is schematically shown in
Fig.~\ref{fig:pll} (top). 

\subsection{Uniform pseudo-magnetic fields and PLLs}

The Hamiltonian is given by Eq.~\eqref{eq:gf}, for which we only retain the strain
component $\vec{\mathcal{A}} \equiv \vec{\mathcal{A}}^3 $,
\begin{gather} \label{eq:hamstrain}
\mathcal{H}(\vec{q}) = \hbar v_F \Gamma_x (q_x +
\mathcal{A}_x \Omega^3)+\hbar v_F \Gamma_y (q_y +
\mathcal{A}_y \Omega^3),
\end{gather}
and $\vec{\mathcal{A}}=\vec{\mathcal{A}}(\vec{x}) $ is taken so
as to describe a constant field. The mathematical structure of the Hamiltonian is
equivalent to that of a real eletromagnetic field and we can use
known techniques to solve it. For completeness we briefly review
the essentials here, leaving more details to Appendix~\ref{app:ll}. We first introduce
dynamical momenta $\hat{\Pi}^\pm_x$ and $\hat{\Pi}^\pm_y$ for each
valley $\nu=\pm$, reflecting the fact the sign of the pseudo-magnetic
field is opposite for the valleys (recall that $\Omega^3=\nu^z$). The
Hamiltonian for each of the valleys reads
\begin{gather} \label{eq:pll}
\mathcal{H}_\pm(\vec{q}) = \pm  v_F  \begin{pmatrix}   &
\hat{\Pi}^\pm_x+i \hat{\Pi}^\pm_y    \\   \hat{\Pi}^\pm_x- i \hat{\Pi}^\pm_y &    \end{pmatrix}.
\end{gather}
and the dynamical momenta obey the commutation relations
\begin{gather}
[\hat{\Pi}^\pm_x, \hat{\Pi}^\pm_y ] = \mp i \frac{\hbar^2}{l^2_B}. 
\end{gather}
These commutation relations can be used to define raising and lowering
operators in the standard way (see App.~\ref{app:ll}), in terms of
which the Hamiltonian takes the form
\begin{gather} \label{eq:hamraislow}
\mathcal{H}_\pm = \pm \begin{pmatrix}   & \sqrt{2}\xi\low_\pm    \\ \sqrt{2}\xi\rais_\pm
  &    \end{pmatrix}.
\end{gather}
Here we have defined $\xi^2 = v^2_F \hbar^2/l_b^2$, and the the
raising and lowering operators obey the commutation relation
$[\low_\pm, \rais_\pm] = \pm 1$. This commutation relation is a key
feature of time-reversal invariant pseudo-magnetic fields, since it
reflects anti-parallel field alignment in the two valleys. The operation of
raising and lowering is interchanged for the two valleys,
which has important consequences for the structure of the
eigenstates. In particular, the eigenstates of the PLL zero modes are
localized on the \emph{same} sublattice, instead of on opposite
sublattices. More specifically one finds
\begin{gather} \label{eq:pllzero}
\ket{\Psi^+_{0}} =\begin{pmatrix} 0 \\ \ket{ \varphi_{0,k}
  }  \end{pmatrix}, \quad \ket{\Psi^-_{0}} =\begin{pmatrix} \ket{ \varphi^*_{0,k}
  } \\ 0 \end{pmatrix}.
\end{gather}
We stress that this implies localization on the same sublattice, given the
choice of basis in Eq.~\eqref{eq:diracspinor}. Time-reversal symmetry is preserved by
counterpropagation of Landau orbitals in the two valleys (i.e., $\ket{ \varphi^*_{0,k}
  } = \ket{ \varphi_{0,-k} }$). The $n=0$ PLL has energy
 $E=0$. Eigenstates corresponding to $n\neq 0 $ PLLs take the form 
\begin{gather} \label{eq:plles}
\ket{\Psi^+_{n\pm}} = \frac{1}{\sqrt{2}}\begin{pmatrix}  \ket{
    \varphi_{n-1,k} } \\ \pm  \ket{ \varphi_{n,k}
  }  \end{pmatrix}, \ket{\Psi^-_{n\pm}} = \frac{1}{\sqrt{2}}\begin{pmatrix}  \ket{
    \varphi^*_{n,k} } \\ \mp \ket{ \varphi^*_{n-1,k} }  \end{pmatrix}
\end{gather}
and they have energies $E_\pm(n) = \pm \sqrt{2\xi^2 n }$ for each
valley.  

\begin{figure}
\includegraphics[width=\columnwidth]{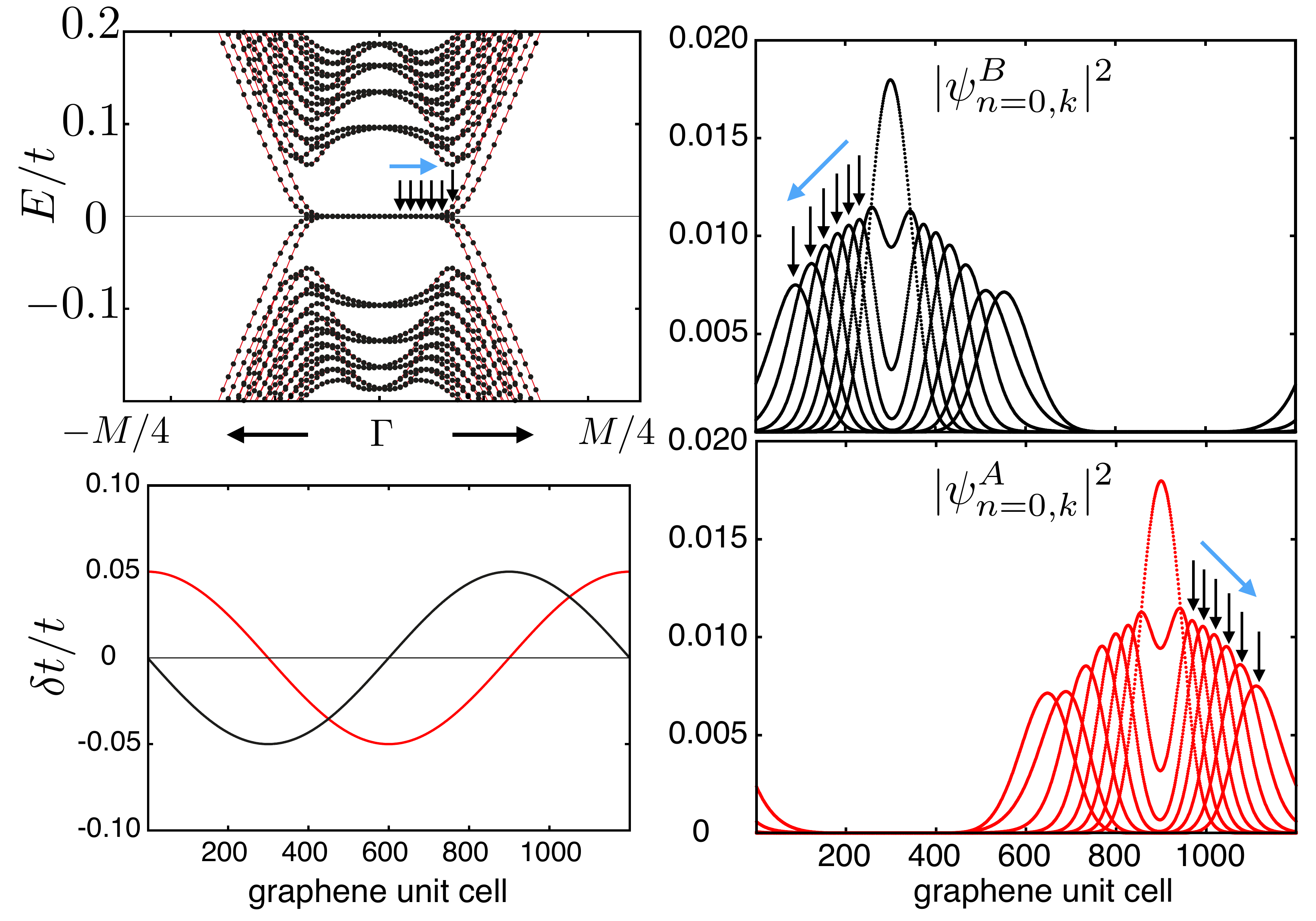}
\caption{\label{fig:spec1} Upper left: Spectrum in the presence of a
  strain-induced periodic pseudo-magnetic field, shown with both Dirac
  points folded onto $\Gamma$. Periodicity of the pseudo-magnetic
  field is $1200$ graphene unit cells, and we used
  $A  =0.05t$. Black arrows indicate for which
$k$ the zero mode eigenstates are plotted in the upper and lower right
panels. In these two panels we show the wave function distribution
$\sim |\psi^{A,B}_{n=0,k}|^2$ of the full zero mode (i.e. $n=0$) subspace over the graphene unit
cells for the $A$-sublattice
(lower right, red) and $B$-sublattice
(upper right, black). Black arrows indicate which $k$ they correspond to in
the upper left plot; the blue arrow indicates in which order. Lower
left: plot of the periodic strain modulation
$\mathcal{A}_x=A\cos (2\pi y/\lambda) \sim
2\delta t_1(y)-\delta t_2(y)-\delta t_3(y)$ (red) and corresponding pseudo-magnetic
field. Note that for clarity we have rescaled the amplitude of the pseudo-magnetic
field to $A$.}
\end{figure}

In order to gain insight into the effect of perturbations on periodic
strain superlattices, we review the effect of such perturbations on the PLL spectra
for uniform pseudo-magnetic field we just presented. Consider the $n=0$ PLL, the eigenstates of which are given in
Eq.~\eqref{eq:pllzero}. Let us first comment on the mass gap terms
$\nu^z\tau^z$ and $\tau^z$. The effect
of the sublattice CDW $\nu^z\tau^z$ and the TRS breaking Haldane term 
$\tau^z$ is reversed as compared to real magnetic fields~\cite{herbut08}. The
sublattce polarized CDW
simply shifts the energy of zero modes but does not break their
degeneracy, whereas the Haldane mass $\tau^z$ energetically splits the zero modes in
a symmetric way. This is an immediate consequence of the sublattice
structure of the PLL zero modes. The two Kekul\'e masses $\nu^x$ and
$\nu^y$ do not affect the zero modes at all, they are neither split
nor shifted, since they are off-diagonal is sublattice space.  

Perturbations that do split the zero modes in a fashion similar to the
Haldane term are charge density waves with tripled unit cell,
i.e., charge density waves that couple the valleys $K_+$ and
$K_-$~\cite{gopa12}. These charge density waves couple to
the Dirac mastrices $\nu^1\tau^1$, $\nu^1\tau^2$, $\nu^2\tau^1$ and $\nu^2\tau^2$. Projecting these into the PLL zero mode
subspace, one finds effective Pauli matrices $\tilde{\tau}^x$ and
$\tilde{\tau}^y$, which anticommute with the Haldane mass
projected into the zero mode space, $\tau^z\rightarrow
\tilde{\tau}^z$. This leads to the counterintuitive situation
of anticommuting masses only one of which is TRS breaking and
nontrivial~\cite{ghaemi13}. We note that these charge density waves with
tripled unit cell correspond to the other gauge field components of
the $SU(2)$ gauge field, i.e. they enter as $\mathcal{A}^{1}_x$,
$\mathcal{A}^{1}_y$, $\mathcal{A}^{2}_x$, and $\mathcal{A}^{2}_y$ in Eq.~\eqref{eq:gfcoupling}. Yet
another perturbation that splits the zeroth PLL is the valley mass,
given by $\nu^z$, making the valleys inequivalent, but acting as the
identity in sublattice space. Its spectral effect is equivalent to
that of the Haldane term, meaning a symmetric splitting of the zero
modes. 

Understanding the spectral effect of these Dirac fermion bilinears on the
zeroth PLL gives a first idea of the ways in which their spontaneous
formation can lower the energy for charge neutral systems. For a
more refined understanding of the energetics it is necessary to consider the effect of
perturbations on higher PLLs (i.e., $n\neq 0 $). For both the CDW term
$\nu^z\tau^z$ and
the Haldane term $\tau^z$ all PLLs with $n\neq 0 $ get pushed up or
down in energy depending on whether they have energies $\pm \sqrt{2\xi^2
  n}$, i.e. positive (negative) solutions get pushed up (down). This
is different for the valley mass $\nu^z$, which pushes all PLLs of
valley $K_+$ up and of valley $K_-$ down, effectively splitting all PLLs,
even the $n\neq 0 $ levels, leaving do degeneracies behind. The charge
density waves with enlarged unit cell (i.e., coupling the valleys)
both split and shift the higher order PLLs, which may be seen
straightforwardly by using perturbation theory up to second
order. Based on these considerations we obtain an intuition for the
spontaneous generation of Dirac fermion bilinears due to interactions,
depending on the location of the Fermi level. 

\subsection{Alternating pseudo-magnetic fields and superlattice PLLs}

The assumption of uniform
pseudo-magnetic field is a useful first step towards understanding the
physics of superlattice PLLs. As a next step we consider the case of
alternating pseudo-magnetic field, which for simplicity we will take
as a periodic arrangement of
regions of positive and negative constant field. This will provide
valuable insight in the case of periodic harmonic pseudo-magnetic
fields. Schematically this field arrangement is shown in the top row of
Fig.~\ref{fig:pll}. Figure~\ref{fig:pll} shows how we can think of the
a periodically alternating field as a strain superlattice with
effective ``two-site'' unit cell (i.e., positive and negative field), reminiscent of an antiferromagnet,
leading to a doubling of the PLL degeneracies. For instance, the
space of zero mode PLLs is doubled, since we have the spatial degeneracy in
addition to valley degeneracy. For each valley there is a zero mode
localized in the positive field region, meaning on the $A$ sublattice,
and a zero mode localized in the negative field region, on the $B$
sublattice. 

The additional degree of freedom originating from the periodicity of the
pseudo-magnetic field gives rise to a richer structure of polarized or ordered
states. Focusing on the PLL zero mode subspace, relevant at charge neutrality, there are multiple
ordered states that lift the degeneracy of the zero mode subspace. Two
of them are shown in Fig.~\ref{fig:pll}. The first is a charge ordered
state, where the zero mode PLLs in one of the two spatial regions are
both occupied, leaving the zero modes in the other region unoccupied. This leads to a redistribution of charge between the two
regions and an associated ferroelectric
polarization along the propagation direction of the superlattice wave
vector. In case of graphene this state is realized by the sublattice
CDW, which energetically discriminates the sublattices. The other
state shown in Fig.~\ref{fig:pll} is the valley-ordered quantum Hall
(or Haldane) state. In such state the zero modes corresponding to an
``up'' pseudo-magnetic field are occupied. Note that this implies an
alternating occupation of valleys $K_\pm$, as shown in
Fig.~\ref{fig:pll}. Therefore, this state may be called \emph{anti-ferro-valley}-ordered. In graphene the valley-ordered quantum Hall state is
realized by the time-reversal symmetry breaking Haldane term. A third
PLL polarized state is obtained by occupying the same valley in each
spatial region. This state also breaks time-reversal, but contrary to
the valley-ordered quantum Hall state the pseudo-magnetic field seen
by the occupied PLLs alternates. The inversion of PLL occupation in
one of the two regions with respect to the anti-ferro-valley-ordered state suggest the name \emph{ferro-valley}-ordered. In graphene such state is realized by
the valley mass term.


\begin{figure*}
\includegraphics[width=0.9\textwidth]{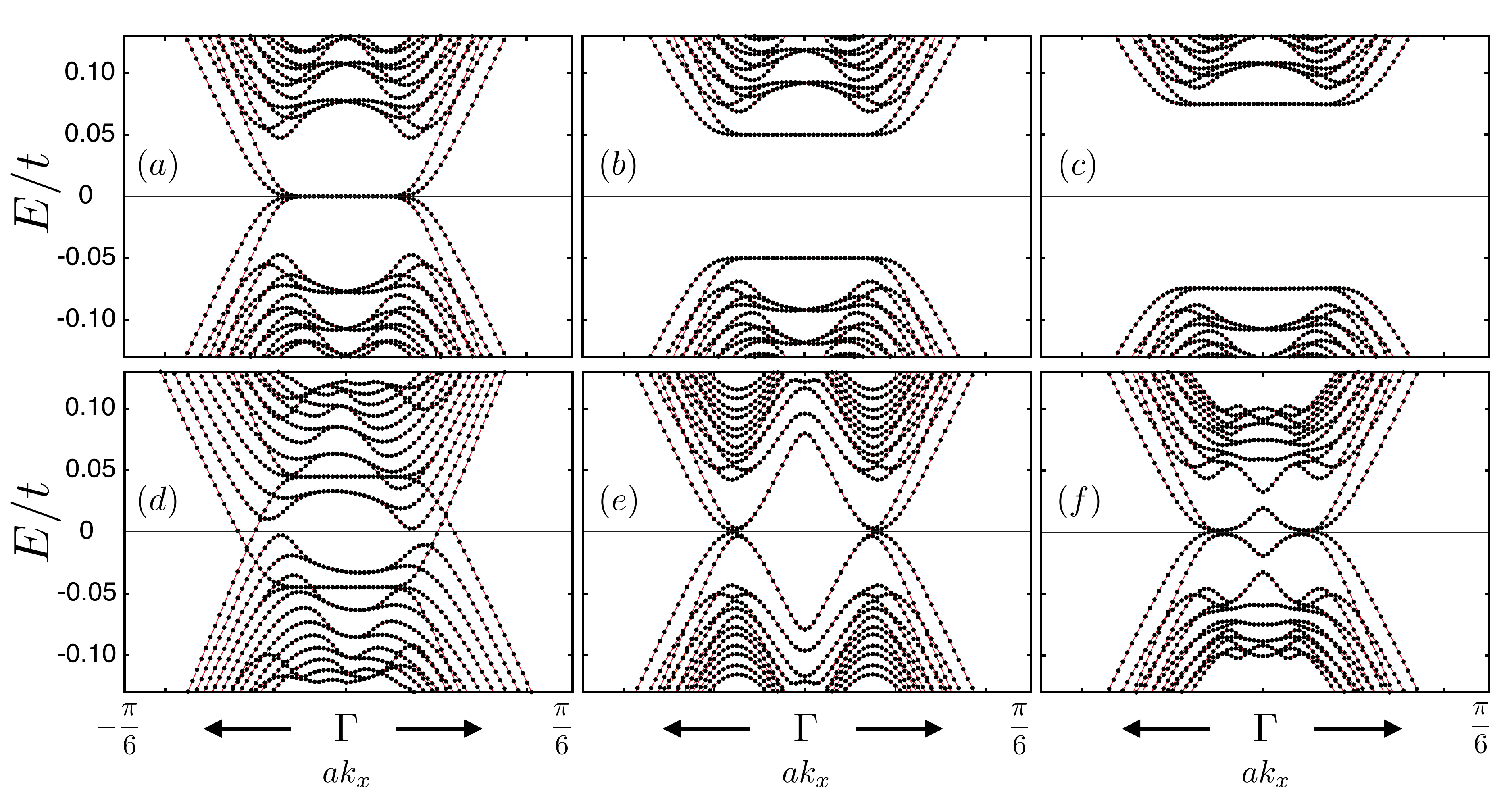}
\caption{\label{fig:specgaps} Spectra of graphene in the presence of
  periodic strain and in the presence of additional perturbations
  splitting or lifting PLL energies. Spectra are obtained for strain
  superlattice unit cells containing $\lambda=600$ graphene unit
  cells and $A = 0.08t$. (a) Free graphene (b) CDW (c)
  Haldane term (d) valley mass term (e) CDW$_1$ with tripled unit cell (f)
  CDW$_2$ with tripled unit cell. }
\end{figure*}

We now establish a connection between the simplified description of alternating
pseudo-magnetic fields in terms of continuum PLLs, and the periodic
strain (super)lattice model introduced in the previous
section. In order to do so we take the unidirectional periodic strain profile
compatible with an enlarged six-site unit cell (i.e., Dirac valleys folded onto
$\Gamma$) defined in Eq.~\eqref{eq:Kmod}. Solving the tight-binding Hamiltonian in the presence of
the strain superlattice yields the spectrum shown in
Fig.~\ref{fig:spec1} (upper left). The connection is made by
interpreting this
spectrum in terms of PLLs. 

Let us first study the wave function support of the zero
energy solutions and compare that to the zeroth PLL. The right column
of Fig.~\ref{fig:spec1} shows the wave function support
$|\psi^{A,B}_{n=0,k}|^2$ of the wave functions $\psi_{n=0,k}$
corresponding to zero energy solutions ($n=0$) labeled by $k$ ($k$
should be identified with $k_y$). Black
arrows explicitly indicate which $k$ corresponds to which
$|\psi^{A,B}_{n=0,k}|^2$ profile. Since there are two valleys and the
strain superlattice unit cell consists of two distinct regions with
opposite pseudo-magnetic field, there is a fourfold degeneracy at each
$k$. This is reflected in Fig.~\ref{fig:spec1} where two pseudo-magnetic Landau-like
orbitals are localized on the $A$-sublattice (lower right) and two
localized on the $B$-sublattice. The wave function support clearly
shows the spatial separation of solutions living on distinct
sublattices. In the region of positive field (see Fig.~\ref{fig:spec1}
lower left) ``zero modes'' are localized on the $A$-sublattice, and
on the $B$-sublattice in the region of negative field. In addition, we
observe that for $k$ moving away from $\Gamma$ (following the blue arrow)
the two Landau orbitals in each region move away from the position of
maximum field towards the position of vanishing field. In particular,
they move away in opposite directions, which is a direct consequence
of their different valley index. As the two valleys effective see
opposite fields, and the spatial position of a Landau orbital is given
by $x \propto \text{sgn}(\mathcal{B})k_y$, the Landau orbitals are expected to
spatially move in opposite directions. With increasing $k$ ($\sim k_y$)
Landau orbitals of different spatial regions and the same valley start
to overlap, eventually leading to the splitting observed in spectrum
indicated by the most right black arrow in the upper left panel of
Fig.~\ref{fig:spec1}. Hence, at the junctions between regions of
positive and negative field, the Landau orbitals acquire a dispersion
and form a series of snake states~\cite{tang14}.

Next, we ask what the spectral effect is of the
perturbations that are expected to split degeneracies, in particular
the flat band degeneracies as depicted in Fig.~\ref{fig:pll}. To this end, we solve the strain superlattice Hamiltonian in the
presence of various
perturbations, which for the moment we take to be spatially uniform,
i.e., not follow the superlattice envelope, in accordance with the schematic picture of
alternating regions of constant field (Fig.~\ref{fig:pll}). Figure~\ref{fig:specgaps} shows the low-energy spectrum
in the vicinity of $\Gamma$. The unperturbed case
corresponding to $\hat{H}_{0} + \hat{H}_{\text{strain}}$ is shown in
Fig.~\ref{fig:specgaps}(a) for reference. The sublattice charge
ordered state and (anti-ferro-))valley-ordered quantum Hall state are shown in Fig.~\ref{fig:specgaps}(b) and
Fig.~\ref{fig:specgaps}(c), respectively. Both open up a full gap,
splitting the zero mode subspace and shifting the higher PLLs, as
expected. Figure~\ref{fig:specgaps}(d) shows the effect of a valley
mass term, i.e., the ferro-valley-ordered, on the low-energy spectrum. Whereas in each
valley degeneracies are preserved, the valleys are
split, as expected. As a consequence, the spectrum is not gapped and the Fermi level
crosses the propagating snake states associated with the flat band
PLLs~\cite{tang14}. 


Figure~\ref{fig:specgaps}(e) and~\ref{fig:specgaps}(f)
show the spectrum obtained in the presence of valley-coupling CDWs,
which we have discussed split the continuum $n=0$ PLL in a way similar to the
Haldane term. We observe that in case of a strain superlattice, the
CDWs do not lead to a full gap, but only lift the degeneracy of the
flat band states localized at the
position where the
pseudo-magnetic field has its extrema. The degeneracy is not lifted in
the vicinity of the nodes of the periodic pseudo-magnetic field. The absence of a full gap in
case of periodic strain can be understood by considering the spectral
effect of the valley-coupling CDWs in case of zero pseudo-magnetic
field. The valley-coupling CDWs do not open up a gap in that case, but
only shift the Dirac points. Hence, at the nodes of the
pseudo-magnetic field one expects the absence of a gap. Note also that the
spectrum numerically obtained numerically shows both split and
shifted higher ($n \neq 0$)
PLLs.

From this analysis we conclude that in the presence of the
strain superlattice, the low-energy electronic structure can be
approximated by sets of PLLs for the two distinct regions of the
superlattice unit cell. In addition, based on their effect on the
degenerate
low-energy PLLs, we expect the charge-ordered
and anti-ferro-valley-ordered states to be the dominant instabilities
in the presence of the pseudo-magnetic field superlattice. 

We end this section with two remarks. First, we
note that the analysis presented here
is based on the the assumption of a strain-induced pseudo-magnetic
field, i.e.,  $\vec{\mathcal{A}}  = \vec{\mathcal{A}}^3$ in Eq.~\eqref{eq:gfcoupling}. As
mentioned in Section~\ref{sec:2ddirac}, a unitary matrix can rotate to
another component, e.g. $\vec{\mathcal{A}}^1$ or
$\vec{\mathcal{A}}^2$. This is does not change the (low-energy) spectrum, but it
does change the nature of the eigenstates. In addition, it also changes the
nature of the perturbations, since the unitary matrix rotates within the
space of masses represented by $\vec{\Gamma}$ as well. In particular, this implies that a
sublattice polarized term $\nu^z\tau^z$ will be rotated into one of the Kekule
terms. Interestingly, the Haldane term is a scalar under these unitary
rotations and hence is invariant. To summarize, the analysis of this
section still
applies, but in a rotated basis. 

The second remark concerns the applicability of our analysis to TCI
surface states. The arguments put forward in the present section
build on the specific example of graphene PLL physics. They remain valid
in the context of TCI surface states. Most importantly, in the
presence of a uniform pseudo-magnetic field, the TCI surface
$n=0$ PLLs are localized on the TCI pseudospin degree of freedom in such a
way that the time-reversal invariant ferroelectric distortion of the
crystal lattice only shifts them, whereas a time-reversal breaking Zeeman-type spin-coupling
splits them in energy. As a result, there is a one-to-one correspondence between the
valley-ordered quantum Hall state and charge-ordered state in graphene, and
Zeeman term and ferroelectric distortion of TCI surface states.




\section{Interacting electrons in a strain superlattice}

In order to systematically investigate the patterns of symmetry
breaking and PLL splitting, as a consequence of interacting flat band
electron, we have studied an interacting Hamiltonian on the graphene
honeycomb lattice and
performed extensive self-consistent Hartree Fock calculations. We
report the results in this section. 

Based on the intuitive picture of the PLLs in the two spatially separated
regions we anticipate both the formation of a charge-ordered state
with ferroelectric polarization,
corresponding to a redistribution of charge between the regions of
positive and negative pseudo-magnetic field, and the formation of a
valley-ordered quantum Hall ground state. Interactions which may lead to the formation of 
these states in a graphene lattice model are the nearest neighbor (NN) and next-nearest neighbor
(NNN) density-density interactions~\cite{raghu08}, respectively, as will be
demonstrated below. We therefore
consider the interacting Hamiltonian $\hat{H} = \hat{H}_0+
\hat{H}_{\vec{\mathcal{A}}} +  \hat{H}_{\text{int}} $ where the interacting part
$\hat{H}_{\text{int}}$ is given by
\begin{gather} \label{eq:intham}
\hat{H}_{\text{int}}  = V_1 \sum_{\langle rr' \rangle} \hat{n}_r
\hat{n}_{r'} + V_2 \sum_{\langle\langle rr' \rangle\rangle} \hat{n}_r
\hat{n}_{r'}.
\end{gather}
Here $V_1$ is the NN interaction strength and $V_2$ the NNN
interaction strength, and $\hat{n}_r$ is the number operator at site $r$. The sums over $\langle rr' \rangle$ and $\langle\langle rr'
\rangle\rangle$ are over NN and NNN, respectively. 

\begin{figure}
\includegraphics[width=0.9\columnwidth]{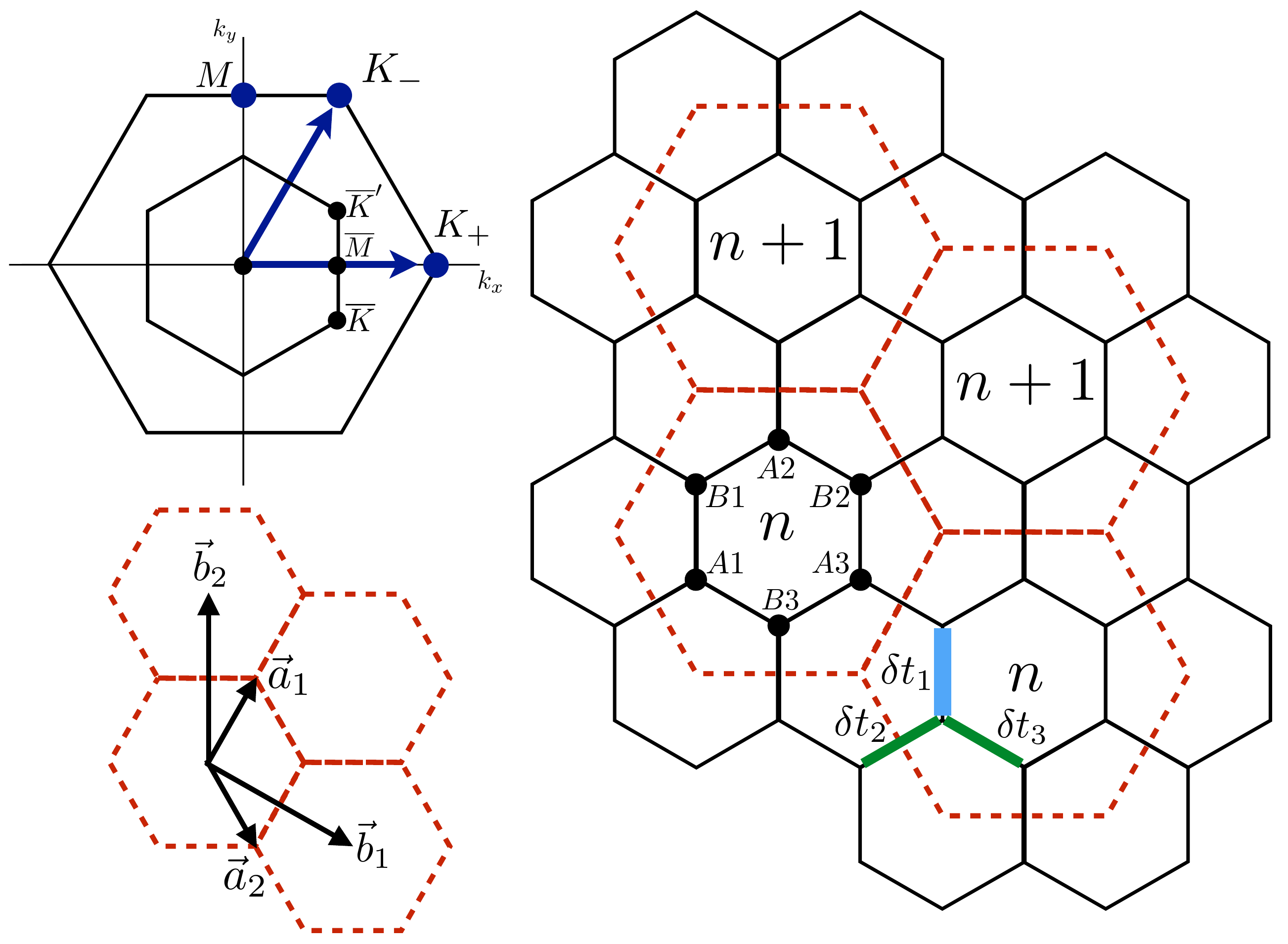}
\caption{\label{fig:lattice} Upper left: Schematic representation of
  the graphene Brillouin zone (outer black hexagon) and the folded
  Brillouin zone corresponding to the tripled unit cell (inner black
  hexagon). Dirac points of pristine graphene are located at $K$ and
  $K'$, which are folded onto $\Gamma$ when tripling the unit
  cell. The vectors connecting $\Gamma$ to $K$ and $K'$, indicated by
  blue arrows, are reciprocal lattice vectors of the enlarged lattice
  vectors. Lower left: Wigner-Seitz sells containing three elementary
  graphene unit cells, with lattice vectors $\vec{a}_1$ and
  $\vec{a}_2$. Right: graphene lattice (black hexagons) and tripled
  unit cells (red dashed hexagons), including labeling of sites. $Ai$
  and $Bi$ label the six sites within the enlarged unit cell;
  $n$ and $n+1$ label the position in the strain-induced
  superlattice. $\delta t_n$ are the hopping amplitude changes due to strain. }
\end{figure}

The spatially modulated strain is implemented in the way described
above, meaning that we take $\hat{H}_{\vec{\mathcal{A}}}$ to contain $\vec{\mathcal{A}}
= (\mathcal{A}_x, 0)$, where the $\mathcal{A}_x$ component of the strain-induced gauge
field originates from hopping amplitude modulations $\mathcal{A}_x \sim 2\delta
t_1-\delta t_2-\delta  t_3 $, which are given a spatial dependence $\delta t_n \rightarrow
 \delta t_n (\vec{r})$. A schematic representation of the way we set
 up the calculations is given in Fig.~\ref{fig:lattice}, which is a
 generalization of the approach of Ref.~\onlinecite{grushin13}
 to the case of strain superlattices (more details
 in Appendix~\ref{app:hf}). In order to
 allow for charge density waves that couple the Dirac points $K_+$ and
 $K_-$ we work with a tripled ($6$-site) unit cell (red
 dashed hexagons in Fig. ~\ref{fig:lattice})~\cite{grushin13}. The corresponding
 lattice vectors are $\vec{b}_1=2\vec{a}_2+\vec{a}_1$ and
 $\vec{b}_2=\vec{a}_1-\vec{a}_2$ in terms of the graphene lattice
 vectors $\vec{a}_i$. We take the strain-induced gauge field to be  
\begin{gather}
\mathcal{A}_x(\vec{r}) = A \cos( \frac{\vec{K}'\cdot \vec{r}}{\lambda/3} )
\end{gather}
where $\vec{r}$ denotes the position of an elementary graphene unit
cell, $\vec{K}$ is the $K_-$ wave vector as shown in
Fig.~\ref{fig:lattice} and $\lambda$ equals the number of graphene
unit cells in the
periodic strain superlattice unit cell. The number of $6$-site unit
cells in the superlattice unit cell is therefore $\lambda/3$. The lattice vectors
of the superlattice are $\vec{b}_1$ and $\lambda\vec{b}_2$. We stress
that we consider a fully periodic system with wavelength $\sim \lambda$
set by the periodic strain superlattice. 

Spinless electrons on the graphene honeycomb lattice serve as a model
system for strain-induced PLLs. We therefore first present Hartree-Fock
results for spinless electrons and then briefly comment on the spin
degree of freedom, recalling that the case of TCI surface states is
similar to spinless graphene electrons.

\subsection{Spinless electrons}

Within the framework of
standard Hartree-Fock theory we decouple the interactions both in
diagonal (Hartree) and off-diagonal (Fock) channels. The diagonal (or
charge density) order parameter at a given site in the superlattice
unit cell is labeled by $(\alpha,i,l)$, and is defined as 
\begin{gather}
\rho_{\alpha i}(l) = \frac{1}{N}\sum_{\vec{k}}\order{\crea{\alpha i}{l,\vec{k}} \anni{\alpha i}{l,\vec{k}} },
\end{gather}
where $\alpha = A,B$; $i=1,2,3$; $l$ labels the $6$-site cell in
the superlattice cell, and $N$ is the total number of
superlattice unit cells. We define $\rho_{\alpha }(l)$ as the average over $i$ within
cell $l$, 
\begin{gather}
\rho_{\alpha }(l) =\frac{1}{3}\sum_{i=1}^3 \rho_{\alpha i}(l).
\end{gather}
Based on previous considerations we are interested in possible local
sublattice imbalances $\Delta_{\rho-}(l) $ expressed as
\begin{gather}   \label{eq:cdwop}
\Delta_{\rho-}(l) =   \rho_{A }(l)-  \rho_{B }(l),
\end{gather}
and possible ferroelectric redistribution of charges between regions
of positive and negative pseudomagnetic field, which can be expressed as
\begin{gather}  \label{eq:ferroop} 
\Delta_{\rho+}(l) =    \rho_{A }(l)+ \rho_{B }(l)-1.
\end{gather}

In addition to the charge density order parameter, in order to detect
the valley-ordered Haldane state, we study the quantum
Hall (QH) order parameter originating from spontaneously generated
next-nearest neighbor tunneling with complex amplitude. Decoupling in
the off-diagonal Fock channel leads to 
\begin{gather} \label{eq:nnnbond}
\chi^\alpha_{il,jl'}  =  \frac{1}{N}\sum_{\vec{k}} \mathcal{Q}^{\alpha
*}_{il,jl'} (\vec{k})\order{\crea{\alpha i}{l,\vec{k}} \anni{\alpha j}{l',\vec{k}} },
\end{gather}
where the phase factors $\mathcal{Q}^\alpha_{il,jl'} (\vec{k})$ are defined in
Appendix~\ref{app:hf}, and from these we extract the QH order
parameter $\Delta_{\text{QH}}(l)$ as 
\begin{gather} \label{eq:qahop}
\Delta_{\text{QH}}(l) =   \sum_{ijl'\in \text{NNN}(l)} \, \eta_\alpha
\text{Im}\;   \chi^\alpha_{il,jl'} .
\end{gather}
The sum is over all NNN pairs which belong to cell $l$ and
$\eta_{A} =-\eta_{B}= 1$ since in the QH state fluxes on the $A$ and $B$
sublattices have opposite sign.  

\begin{figure}
\includegraphics[width=\columnwidth]{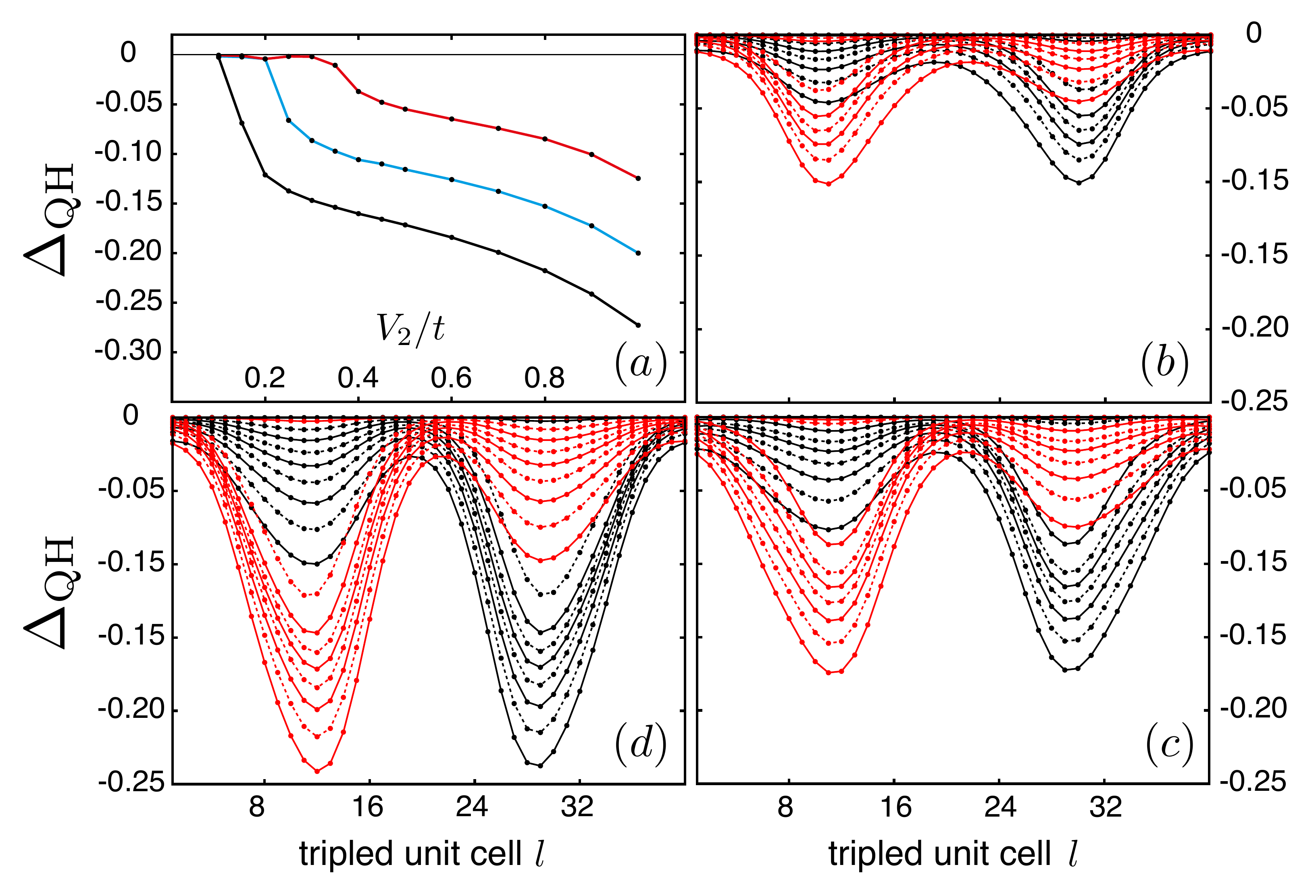}
\caption{\label{fig:qah} Panel (a) shows the maxima of QH order parameter
  $\Delta_{\text{QH}}$ as function of NNN interaction $V_2$ for
  various values of the pseudo-magnetic field amplitude: $A = 0.10$
  (red), $A = 0.15$ (blue), $A = 0.20$ (black). Panels (b)-(c)
  show the QH order parameter as function of tripled unit cell
  $\Delta_{\text{QH}}(l)$ ($l$ labeling tripled unit cell) for the
  same values of $A$ [ (b) $A = 0.10$; (c) $A = 0.15$; (d) $A
  = 0.20$]. Curves are shown for
  $V_2=0.1,0.2,0.3,0.4,0.5,0.6,0.7,0.8,0.9,1.0$, in descending order,
  i.e. bottom most curves $V_2=1.0$ and top most curves $V_2=0.1$. Black and red
  curves correspond to $A$ and $B$ sublattice, respectively. $V_1=0$
  in all cases. }
\end{figure}

The results of self-consistent HF calculations we present here were performed for spinless electrons on lattices of size $N =
16 \times 16$ and $\lambda = 3 \times 40 = 120$. In the following we
map out the phase diagram as function of $V_1$ and $V_2$ by discussing
the results for three specific regimes separately. First we will focus
on $(V_1=0,V_2\neq 0)$ to show that the QH state is stabilized for
the smallest values of $V_2$ as a result of the strain induced flat PLL-like
bands. Second, we will look at the case $(V_1\neq 0,V_2= 0)$ to show
that a NN interaction will induce the ferroelectic charge ordered state. Third and last we will focus on selected cases of $(V_1\neq
0,V_2\neq 0)$ to show that the ferroelectric charge density wave is
strongly suppressed as compared to the valley-ordered QH state, precisely due to the
localization of the low-energy states flat band states on a single
sublattice. 

Figure~\ref{fig:qah} shows the HF results for various values of NNN
interaction $V_2$ while keeping $V_1=0$. Panels (b)-(c) show the QH order
parameter $\Delta_{\text{QH}}(l)$ defined in Eq.~\eqref{eq:qahop} as
function of then tripled cell index $l$. Black and red curves correspond to $A$
and $B$ sublattice, respectively, and the strength of the tunneling
amplitude variation $A$ is $A = 0.10$ (b),
$A = 0.15$ (c), and $A = 0.20$ (c). It is
apparent from these figures that
the QH order parameter follows the profile of the effective
pseudo-magnetic field. In the spatial region where the flat band
states are localized on the $A$ sublattice, the QH order parameter
develops predominantly on the $A$ sublattice, and vice versa for the
$B$ sublattice region. Ordinarily, in the honeycomb lattice QH state, the spontaneously induced magnetic fluxes are opposite on the
$A$ and $B$ sublattice, averaging to zero over an elementary unit
cell~\cite{haldane88}. In the present case the localization of flat band states on a
single sublattice leads to finite fluxes in regions of positive and
negative pseudo-magnetic field, which average to zero only over the
larger strain superlattice unit cell. 

In addition to the locking of the QH order parameter
$\Delta_{\text{QH}}(l)$ to the sublattice structure of the flat band
zero modes, we observe that the strain-induced reorganization of
the low-energy electronic structure into PLLs, fundamentally changes
the impact of interactions. Figure~\ref{fig:qah}(a) shows the
dependence of the QH order $\Delta_{\text{QH}}(l_{\text{max}})$,
where $l_{\text{max}}$ is the cell index where the QH order is
strongest, on the strength of the interaction $V_2$ for various values
of the pseudo-magnetic field strength (i.e.,
$A$). Whereas for the unstrained honeycomb lattice a
finite interaction $V_2 \sim 1.3$ is needed to stabilize the QH state
(within HF
theory)~\cite{raghu08,weeks10,grushin13} due
to the vanishing density of states at half filling, here we find that
the QH state is induced already for small interactions, in particular in
regions where the effective field is strongest. For perfectly flat
bands such as PLLs one expects the interaction-induced order to scale
linearly with interaction for weak coupling
~\cite{kopnin11,uchoa13}. This is reflected in
Fig.~\ref{fig:qah}(a) which shows that for increasing pseudo-magnetic
field, setting the PLL band flatness, the QH is more robust and its
dependence on the interaction is approximately linear, with deviations
at very small values of $V_2$. 

\begin{figure}
\includegraphics[width=\columnwidth]{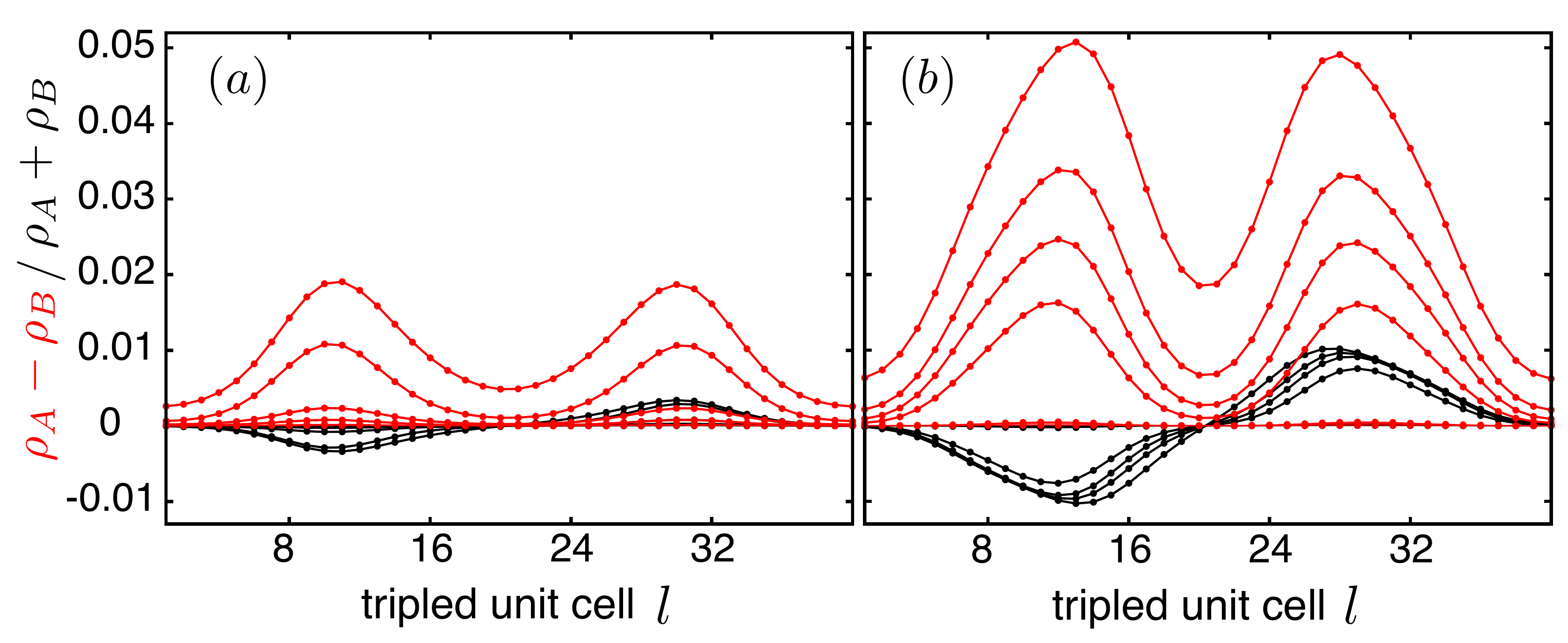}
\caption{\label{fig:cdw} (a) Plots of the charge density wave order
  parameter $\rho_A(l)-\rho_B(l)$ (red) and total charge redistribution
  $\rho_A(l)+\rho_B(l)-1$ (black) for values of the NN interaction
  $V_1=0.2,0.4,0.5,0.6,0.7,0.8$, in both cases plotted in ascending order (top curve
  $V_1=0.8$, bottom $V_1=0.2$); $A=0.10$. (b) Same is (a) but with
  $A=0.20$. }
\end{figure}

Next, we turn to the case of finite NN interaction $V_1$ while keeping
$V_2=0$. In the absence of strain the unfrustrated NN interaction will
favor a CDW characterized by translational symmetry preserving
sublattice charge
imbalance~\cite{raghu08,weeks10,grushin13}. On
the contrary, in
the presence of strain, adopting the PLL picture for the low-energy
electrons and focusing on the
zeroth PLL, the effect of the NN interaction is expected to be suppressed, as the zeroth PLL states live
exclusively on one sublattice. Nevertheless, since the NN has the
potential to cause charge asymmetry between the sublattices (i.e., the
regions of positive and negative pseudo-magnetic field), and
higher PLLs may be relevant to the energetics, we anticipate the system to develop a charge density wave of ferroelectric
type (i.e., $\Delta_{\rho+}$), with excess charge in regions where the pseudo-magnetic field is
positive, and defect charge where it is negatice, or vice
versa. In addition, in the previous section we observed that under the
assumption of a uniform CDW the strain-induced PLL spectrum is gapped out. 

In Fig.~\ref{fig:cdw} we show results for both $\Delta_{\rho+}(l)$
and $\Delta_{\rho-}(l)$, defined in Eqs.~\eqref{eq:ferroop} and
\eqref{eq:cdwop}, for different values of the interaction $V_1$ and
strain $A$. As expected, we find the CDW order parameter
$\Delta_{\rho-}(l)$ (shown in red)
to become finite in regions where the pseudo-magnetic field is
strongest, but to have the same sign in both positive and negative
regions. The sign of concomittant ferroelectric polarization depends
on the sign of the pseudo-magnetic field. In the region where the
field is positive the flat band states are localized on the $A$
sublattice and pushing them down in energy, signaled by positive
$\Delta_{\rho-}(l)$, leads to excess charge in that region at the expense
of charge in the region of negative field. At the same time we observe
that the FP is more pronounced for stronger strains, and that it is
very weak for small interaction. The latter may be attributed to the
fact that NN interactions have no effect in the zeroth PLL. 

We proceed to consider the case of both finite $V_1$ and $V_2$. We
have seen that both of these interaction individually favor different
gapped ground states. At the same time we argued that as a consequence
of the different nature of these interactions, i.e. $V_1$ being
inter-sublattice and $V_2$ being intra-sublattice, they have different
impact on the low-energy flat band electrons. Due to the spatial
separation of states localized on different sublattices, it is
expected that the effect of $V_1$ is suppressed. One therefore expects
the QH state to survive even for $V_2<V_1$, which corresponds to the
physically relevant regime. 

\begin{figure}
\includegraphics[width=0.8\columnwidth]{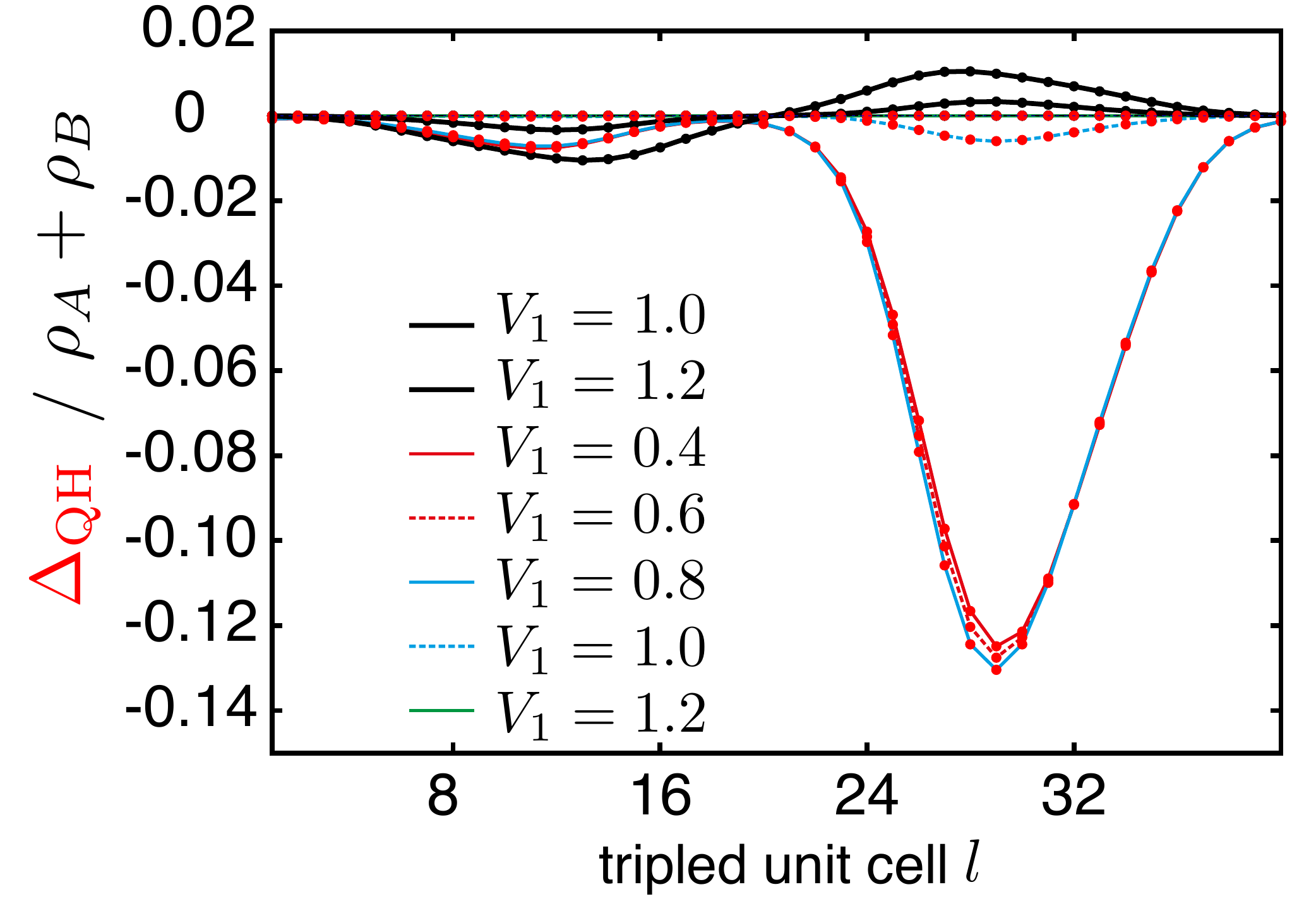}
\caption{\label{fig:qahcdw} Plot of both the QH order parameter
  $\Delta_{\text{QH}}(l)$ (only $A$-sublattice shown; red dot curves) and total
  charge redistribution $\rho_A(l)+\rho_B(l)-1$ (black dot curves) as function of tripled
  unit cell index $n$ for $A=0.20$ and $V_2=0.20$. Different curves
  correpond to different values of $V_1$, explicitly labeled for clarity.}
\end{figure}

In Fig.~\ref{fig:qahcdw} we present results for various $V_1$ with
finite $V_2=0.2 $ and for $A=0.20$. We plot both the
ferroelectric order parameter $\Delta_{\rho+}(l)$ and the QH order
parameter $\Delta_{\text{QH}}(l)$ for the $A$ sublattice. The key
observation is that even for small $V_2=0.2 $ the QH survives up to
NN interactions $V_1 \sim 1.0$. The therefore conclude that the effect
of interactions on periodic strain induced flat bands follows from
their PLL character. In particular, the sublattice structure of the low-energy
flat bands is the decisive factor in determining which order is
spontaneously generated by interactions. 

\subsection{Remarks on spinful electrons}

We close this section with a number of remarks on the electron
spin. The numerical calculations have been performed for spinless
electrons in graphene. Taking spin into account gives rise to a
richer structure of polarized states, specifically the ferromagnetically
(FM) and anti-ferromagnetically (AFM) polarized states should then be considered, in
addition to the Quantum Spin Hall polarized state. Moreover, the
argument for the suppression of the NN interactions does not apply to
the onsite Hubbard interaction, $ \hat{H}_{U}  = U \sum_{r} \hat{n}_{\up r}
\hat{n}_{\down r} $, which must be included in the interacting
Hamiltonian. 

The results for spinless electrons in graphene do, however, directly
apply to TCI surface, which do not have an additional degenerate
spin degree of freedom. Instead, as a consequence of spin-orbit
coupling, spin is already part of the low-energy Dirac
structure. In particular, our results imply that on the surface of a
TCI and in the presence of periodic strain, interactions will lead to
the formation of the QH state. 


As a first step towards understanding the polarization of spinful strain
superlattice-induced PLLs in graphene, we have performed numerical Hartree-Fock
calculations with an interacting Hamiltonian given by $H_U$. We find that the mean-field ground is a \emph{superlattice
anti-ferromagnet}. The superlattice anti-ferromagnet exhibits
anti-ferromagnetic order, as expected on a bipartite honeycomb
lattice, however, since the flat band states are localized on one
sublattice only, in each of the two regions of the strain
superlattice an effective magnetization develops. Hence, as a consequence of
the particular structure of the zeroth PLL states, the
anti-ferromagnetic order is transferred to the superlattice. 

A similar result was reported in Ref.~\onlinecite{roy14b}, which found
anti-ferromagnetic order induced by non-periodic strain, where the
bulk and the sample boundary have an effective but opposite magnetization.




\section{Discussion and conclusion}

We have shown that in the presence of a strain superlattice, a
periodic modulation of elastic lattice deformations, a system of
low-energy Dirac electrons exhibits a fourfold degenerate zero energy
flat band, reminiscent of a zeroth PLL. The PLL structure originates from
the pseudo-magnetic field, generated by nonuniform strain. The strain superlattice unit
cell consists of two spatially distinct regions, one in which
electrons see a positive pseudo-magnetic field and one in which they
see a negative field. The single-particle states of the
degenerate flat band have a special and important localization property: in each of
the two regions their wavefunction has support only on one of the Dirac pseudospin
species. 

Periodic pseudo-magnetic fields can occur both in graphene and on the
surface of a TCI, which hosts pairs of Dirac fermions at opposite momenta related by
time-reversal symmetry. The important fact that Dirac fermions are 
unpinned to time-reversal-invariant momenta in the BZ allow for
pseudo-gauge field under time-reversal invariant perturbations such as strain. 

Interactions between electrons in continuum PLLs are expected to lead to the formation of polarized
states splitting the degeneracies of PLLs. We have investigated
PLL polarization for the case of lattice PLLs corresponding to
periodic pseudo-magnetic fields. Two polarized states were shown to
fully lift the zero energy flat band degeneracy while at the same time
pushing all occupied (unoccupied) lattice PLLs down (up). The first
state is the sublattice polarized charge-ordered state, which can be pictured
as a spatially polarized state with all PLLs of the positive (or
negative) field region occupied. The second is the anti-ferro-valley
ordered state,
or spontaneous quantum Hall state, for which all PLL states effectively seeing a positive (or
negative) field are occupied, implying time-reversal symmetry
breaking. We found that other polarized states, even though they have similar
characteristics in the continuum, do not fully lift the lattice flat
band PLL degeneracies. 

Using self-consistent Hartree-Fock calculations we have studied an
interacting honeycom lattice model with periodic strain-induced
lattice PLLs. Our results demonstrate that the strain-induced
reconstruction of low-energy elecronic structure, in particular the
presence of a
zero energy flat band, determines the impact of interactions. Three
key results highlight this conclusion. First, the mean-field order
parameters clearly reflect the periodicity of the pseudo-magnetic
field, showing that the amplitude of the order parameter is tied to
the strength of pseudo-magnetic field. 

Second, as a
consequence of the characteristic wavefunction support of the flat band
single-particle states, the effect of interactions that favor
time-reversal invariant pseudo-spin order is suppressed.
This effectively enhances interactions that favor time-reversal
symmetry breaking valley order with associated spontaneous
quantum Hall effect. We have established this result in the context of the
graphene lattice model, where the pseudo-spin
corresponds to the sublattice degree of freedom. The NN interactions
are inter-sublattice interactions, and therefore supressed, as the
flat band single-particle states have support on one of the
sublattices only, in each positive or negative field region.
The result, however, is general and applies equally to TCI surface
states. The physical interpretation of pseudo-spin and valley are
different, as explained in Sec.~\ref{sec:2ddirac}, yet the effect of
interactions favoring time-reversal symmetry breaking remains
strongly enhanced.

Third, the valley-ordered spontaneous quantum Hall state
already occurs for small interactions when the PLLs are
well-developed. The PLL structure is a way to significantly enhance density of states
near the charge neutral point. 
We conclude that in the presence of periodic strain and interactions, a
system of unpinned Dirac electrons has a generic instability towards a
spontaneous quantum Hall phase. 

The emphasis of the numerical calculations we report, has been on
spinless electrons in graphene. In TCI surface states, however, no additional spin degeneracy is
present. Due to spin-orbit coupling the electron spin is an intrisinc
part of the low-energy Dirac structure. In particular, this implies
that there are no purely spin-polarized phases, such as the global anti-ferromagnet, competing with the
spontaneous quantum Hall phase, favoring the latter as ground state.



Whereas uniform strain-induced pseudo-magnetic fields suffer from
implementation limitations, particularly beyond the nano-scale,
periodic strain can potentially be realized in macroscopic sample
sizes. In fact, such periodic strain fields and induced
pseudo-magnetic fields were demonstrated in TCI heterostructure~\cite{palatnik81,springholz01}, making TCI surface states the prime candidate to exhibit
spontaneous formation of nontrivial electronic states.

\begin{acknowledgements}
We thank Evelyn Tang, Vlad Kozii and Sarang Gopalakrishnan for interesting discussions. J.V. acknowledges support
from the Netherlands Organization for Scientific Research (NWO). L.F. is supported by  David and Lucile Packard Foundation. 
\end{acknowledgements}

\appendix

\section{Landau levels for Dirac electrons\label{app:ll}}

For the purpose of being selfcontained we collect some standard
results of Dirac fermions in a constant magnetic field in this
Appendix. These may be directly applied to the case of time-reversal
invariant pseudo-magnetic fields, bearing in mind the key
characteristic of opposite sign of the pseudo-magnetic field in the
two valleys. 

In the presence of a magnetic field we define the dynamical momenta
using the Peierls substitution
\begin{gather*}
\hat{p}_\alpha \; \rightarrow \; \hat{\Pi}_\alpha = \hat{p}_\alpha - e A_\alpha(\hat{r})  = -i\hbar \partial_\alpha +|e|A_\alpha(\hat{r}),
\end{gather*}
where $\hat{p}_\alpha$ is the momentum operator and $\alpha=x,y$ (we
restrict the description to two dimensions). $\hat{r}_\alpha$ are the
position operators obeying $[\hat{r}_\alpha, \hat{p}_\beta] = i\hbar $
and $A_\alpha(\hat{r})$ is the electromagnetic gauge field. In a
magnetic field the momentum operators $\hat{\Pi}_\alpha$ do not commute but instead obey the canonical commutation relation 
\begin{gather*}
[\hat{\Pi}_\alpha, \hat{\Pi}_\beta ]  = |e| [ \hat{p}_\alpha  ,
A_\beta(\hat{r})   ]+|e|[ A_\alpha(\hat{r}) ,\hat{p}_\beta  ]  \\
=-i\hbar |e| F_{\alpha\beta}
\end{gather*}
where $F_{\alpha\beta} = \partial_\alpha A_\beta - \partial_\beta
A_\alpha$ is the field strength. Assuming a uniform field strength in
the $\hat{z} $ direction the magnetic field is given by $B_\lambda =
\epsilon_{\lambda\mu\nu}F_{\mu\nu}/2$, implying that $F_{\mu\nu} =
\epsilon_{\mu\nu\lambda}B_\lambda$. In particular we have for a uniform field $B \equiv B_z$ in the $\hat{z} $ direction
\begin{gather} \label{eq:appcomrel}
[\hat{\Pi}_\alpha, \hat{\Pi}_\beta ] = -i\hbar |e|\text{sgn}(B) B  \epsilon_{\alpha\beta z}. 
\end{gather}
In this expression we have explicitly separated the sign of the
magnetic field from its strength $B=|B|$ so as to make dependencies on
the sign of the field transparent. Defining the fundamental
characteristic length scale in the system, the magnetic length, as
$l_B = \sqrt{\hbar /(|e|B})$, we can write $[\hat{\Pi}_\alpha,
\hat{\Pi}_\beta ] = -i \hbar^2 \text{sgn}(B)\epsilon_{\alpha\beta
  z}/l_B^2$. This implies a canonical commutation relation between the
dynamical momenta and inspires to define creation and annihilation
operators in the usual way as
\begin{eqnarray}
\hat{a}^\dagger &=& \frac{l_b}{\sqrt{2}\hbar} (\hat{\Pi}_x +i \text{sgn}(B)\hat{\Pi}_y), \nonumber \\
 \hat{a} & = & \frac{l_b}{\sqrt{2}\hbar} (\hat{\Pi}_x - i \text{sgn}(B) \hat{\Pi}_y),
\end{eqnarray}
which obey $[\hat{a}, \hat{a}^\dagger] = 1$. Note that the definition
of these operators depends on the sign of the $B$-field, which is a
direct consequence of Eq.~\eqref{eq:appcomrel}. We note in passing
that all of the above did not require specifying a gauge for $A_\alpha$. 

The Landau level spectrum of a Dirac Hamiltonian of the form
\begin{gather*}
 \mathcal{H} =  \hbar v_F (\Gamma_x q_x +\Gamma_y q_y )
\end{gather*}
is then straightforwardly obtained by making the substitution $\hbar q_\mu
\rightarrow \hat{\Pi}_\mu$. Squaring the Hamiltonian yields
\begin{gather*}
 \mathcal{H}^2 =  v^2_F ( \hat{\Pi}^2_x+\hat{\Pi}^2_y) +
 v^2_F[\hat{\Pi}_x, \hat{\Pi}_y]\Gamma_x\Gamma_y  \\
= \frac{v^2_F \hbar^2}{l_b^2} ( 2\rais\hat{a} -1 +\text{sgn}(B) \tau_z)
\end{gather*}
We use that $\rais\hat{a} = n$ for standard oscillator wave functions
$\varphi_n$, i.e. $\rais\hat{a} \varphi_n = n \varphi_n $, $\hat{a}
\varphi_n = \sqrt{n} \varphi_{n-1}$ and $\rais \varphi_n = \sqrt{n+1}
\varphi_{n+1}$. One therefore obtains the Landau level energies  
\begin{gather} \label{eq:llenergy}
E_{\pm}(n) = \pm \sqrt{2\xi^2 n }, \quad n=1,2,\ldots,  \nonumber \\
 \xi^2 \equiv \frac{v^2_F \hbar^2}{l_b^2} 
\end{gather}
Each of the $E_{\pm}(n) $ is two-fold degenerate because of the valley
degree of freedom, in addition to an $N_\phi= A/ 2\pi l_b^2$
degeneracy where $A$ is the area of the system. 

We find the corresponding eigenstates by taking a closer look at the
explicit expression for the Hamiltonian in each valley. Writing
$\mathcal{H}_\nu$ for the Hamiltonian in valley $\nu = \pm$ we have
\begin{gather*}
\mathcal{H}_\nu = \nu \begin{pmatrix}   & \sqrt{2}\xi \low    \\ \sqrt{2}\xi\rais   &    \end{pmatrix}.
\end{gather*}
The eigenstates belonging to the eigenvalues $E_{\pm}(n) $ of
Eq.~\eqref{eq:llenergy} are easily obtained as
\begin{gather} \label{eq:es}
\ket{\Psi_{n\nu\pm}} = \frac{1}{\sqrt{2}}\begin{pmatrix}  \ket{
    \varphi_{n-1,k} } \\ \pm \nu  \ket{ \varphi_{n,k} }  \end{pmatrix},
\end{gather}

In addition to the states $\ket{\Psi_{n\nu\pm}} $ ($n=1,2,\ldots$)
there are also zero mode states $\ket{\Psi_{0\nu}} $, one for each
valley, which have zero energy ($E_0=0$). Inspecting of
Eq.~\eqref{eq:hamraislow} reveals that these states are given in each
of the valleys as 
\begin{gather} \label{eq:eszero}
\ket{\Psi_{0\nu}} =\begin{pmatrix} 0 \\ \ket{ \varphi_{0,k} }  \end{pmatrix}.
\end{gather}
We stress that this implies the zero modes are localized on
\emph{opposite} sublattices for the two valleys, as we had exchanged
sublattices for the $\vec{K}_-$ valley. 

We proceed to consider the effect of symmetry breaking terms on the
Landau level spectrum. Specifically, we consider first the set of
time-reversal symmetry invariant masses $\vec{m}$ which enter the
Hamiltonian as 
\begin{gather*} 
\mathcal{H}_{\vec{m}} = \vec{m}\cdot \vec{\Gamma} = m_1\nu^x+m_2\nu^y+m_3\nu^z\tau^z
\end{gather*}
 For small masses we may use perturbation theory to study the
 splitting or shifting of Landau levels. It turns out however that the
 exact energies can be obtained in the presence of
 $\mathcal{H}_{\vec{m}} $. The energies are found by squaring the
 Hamiltonian $\mathcal{H}=\mathcal{H}_0+\mathcal{H}_{\vec{m}}$, which
gives
\begin{gather*}
 \mathcal{H}^2 =  v^2_F ( \hat{\Pi}^2_x+\hat{\Pi}^2_y) +
 v^2_F[\hat{\Pi}_x, \hat{\Pi}_y]\Gamma_x\Gamma_y + m^2  \\
=  \frac{v^2_F \hbar^2}{l_b^2} ( 2\rais\hat{a} -1 +\text{sgn}(B) \tau_z) +  m^2,
\end{gather*}
where we use the anticommutation relations of the $\Gamma$-matrices. We directly find the energies
\begin{gather}
E_{\pm}(n) = \pm \sqrt{2\xi^2 n   +  m^2}, \quad n=1,2,\ldots
\end{gather}
Expanding the square root $\sqrt{2\xi^2 n}\sqrt{1 +  m^2/2\xi^2 n}$ in
small $m^2/2\xi^2 n$ yields the same result as second order
perturbation theory. 

The Landau level spectrum in the presence of masses can alternatively
obtained by using the relation $[\Omega^i,\Gamma_j] =
2i\epsilon_{ijk}\Gamma_k$ to construct a unitary matrix $U$ which
rotates the vector $\vec{m}$ so that one has $U^\dagger \vec{m}\cdot
\vec{\Gamma} U = m \Gamma_3$, with $m = |\vec{m}|$. Such transformation block diagonalizes
the Hamiltonian and we obtain the energies in a more direct way. In
particular, we can employ the unitary rotation to find the energies of
the zero modes for the case of massive Dirac fermions. In the rotated
basis we can construct zero mode states in the same way as before,
which will have energies 
\begin{gather}
E_{0\nu} = -\nu m
\end{gather}
where $\nu = \pm$ represents the valley degree of freedom. 

We conclude by taking into account the time-reversal symmetry breaking
but chiral symmetry preserving mass $\eta$ entering as
$\mathcal{H}_\eta= \eta \tau^z$. Since such this term is a scalar
under chiral rotations generated by $\Omega^i$ we may consider
$\mathcal{H} = \mathcal{H}_0  +
\mathcal{H}_{\vec{m}}  +\mathcal{H}_\eta $ and use the chiral rotation
$U$ to block diagonalize the Hamiltonian. The total mass term then is
$m\nu^z\tau^z+\eta\tau^z$, which directly leads to the energies
\begin{gather}
E_{\nu\pm}(n) =   \pm \sqrt{2\xi^2 n + (\nu m+\eta)^2} 
\end{gather}
for the Landau levels $n=1,2,\ldots$. The presense of both of these
masses leads to a splitting of Landau levels, whereas the presence of
either only shifts the energies. The presence of a time-reversal
breaking mass breaks particle-hole symmetry in the $n=0$ Landau
level. Specifically, the zero mode energies become
\begin{gather}
E_{0\nu} = -\nu m - \eta.
\end{gather}





\section{Setup of Hartree-Fock calculations\label{app:hf}}

Our Hartree-Fock calculations in the presence of the strain
superlattice and with interacting Hamiltonian~\eqref{eq:intham} follow
the scheme of Ref.~\onlinecite{grushin13}. In particular, we choose
the unit cell of the unstrained lattice such that it contains six honeycomb lattice sites, as show
in Fig.~\ref{fig:lattice} of the main text. This allows
intra-sublattice mean-field structures to form, corresponding to modulation
vectors $K_{\pm}$, which connect the Dirac points of the honeycomb
lattice. In terms of the elementary graphene lattice vectors
\begin{gather*}
\vec{a}_1 = a (1,\sqrt{2})/2, \quad \vec{a}_2 = a (1,-\sqrt{2})/2,
\end{gather*}
the lattice vectors of the six-site unit cell are given by
\begin{gather*}
\vec{b}_1 = 2 \vec{a}_1 + \vec{a}_2, \quad \vec{b}_2 = \vec{a}_1 - \vec{a}_2.
\end{gather*}
They are shown in Fig.~\ref{fig:lattice}, together with the folded
BZ. Figure~\ref{fig:lattice} also shows how, in the presence of the
periodic strain superlattice, the superlattice unit cell is
defined. The superlattice vectors are $\vec{b}_1$ and
$\lambda\vec{b}_2$, in terms of the superlattive wave length $\lambda$.

The electronic Hamiltonian is expressed in terms of the fermion
annihilation (and corresponding creation) operators $\anni{\alpha
  i}{l,\vec{x}} $. Here $\alpha$ labels th sublattice ($A/B$),
$i=1,2,3$ labels the three sites of each sublattice species in the
six-site unit cell, and $l=1,\ldots,\lambda$ lables the (six-site)
unit cells in the superlattice unit cell, and $\vec{x}$ is a position
index for the superlattice unit cell. The Fourier transforms is
defined as 
\begin{gather*}
\anni{\alpha i}{l,\vec{k}} = \frac{1}{\sqrt{N}}\sum_{\vec{x}}
\anni{\alpha i}{l,\vec{x}} e^{-i\vec{x}\cdot \vec{k}} .
\end{gather*}

The interacting Hamiltonian of Eq.~\eqref{eq:intham} consists of two
terms: the NN interaction and the NNN interaction. In momentum space
the NN interaction $\hat{H}_{V_1} $ is given by
\begin{gather}
\hat{H}_{V_1}  = \frac{V_1}{N} \sum_{\vec{k}\vec{k}'\vec{q}} \crea{A
  i}{l,\vec{k}} \anni{A i}{l,\vec{k}-\vec{q}}
\mathcal{X}_{il,jl'}(\vec{q}) \nonumber \\
 \times \crea{B j}{l',\vec{k}'} \anni{B j}{l',\vec{k}'+\vec{q}} , \label{eq:hamv1}
\end{gather}
where repeated indices are summed. The matrix function
$\mathcal{X}(\vec{q})$ connects NNs, and it is convenient
to decompose it in the following way
\begin{gather*}
\mathcal{X}_{il,jl'}(\vec{q}) =
\mathcal{X}^{-}_{ij}(\vec{q})\delta_{l,l'-1} + \mathcal{X}^0_{ij}(\vec{q})\delta_{l,l'}+\mathcal{X}^+_{ij}(\vec{q})\delta_{l,l'+1}.
\end{gather*}
Clearly, $\mathcal{X}^0_{ij}(\vec{q})$ connects sites within the same
(six-site) unit cell, and it is explicitly given by
\begin{gather}
\mathcal{X}^0_{ij}(\vec{q}) = \begin{pmatrix}  1  & 0   &  1   \\ 1
&  1   &   0 \\  e^{i\vec{b}_1\cdot \vec{q}}   &  1  & 1  \end{pmatrix}.
\end{gather}
The functions $\mathcal{X}^{\pm}_{ij}(\vec{q})$ connect sites in
different (six-site) unit cells and each only have a single nonzero
entry. They are given by $\mathcal{X}^+_{12}(\vec{q})  =
e^{-i(\vec{b}_1+\vec{b}_2)\cdot \vec{q}}$ and $\mathcal{X}^{-}_{23}(\vec{q})  =  e^{i\vec{b}_2\cdot \vec{q}}$.

Similarly, the NNN interaction Hamiltonian is given by
\begin{gather}
\hat{H}_{V_2}  = \frac{V_2}{2N} \sum_{\vec{k}\vec{k}'\vec{q}} \crea{\alpha
  i}{l,\vec{k}} \anni{\alpha i}{l,\vec{k}-\vec{q}} \mathcal{Y}^{\alpha}_{il,jl'}(\vet{q}) \nonumber \\
 \times\crea{\alpha j}{l',\vec{k}'} \anni{\alpha
   j}{l',\vec{k}'+\vec{q}}  \label{eq:hamv2}
\end{gather}
The functions $\mathcal{Y}^{\alpha}(\vet{q}) $ connect NNNs on each
sublattice $\alpha$. They are directly obtained from
Ref.~\onlinecite{grushin13}, taking into account the additional
superlattice index $l$ (in the same way as for
$\mathcal{X}(\vec{q})$). 




The quartic interactions of the interacting Hamiltonians,
schematically written as $\hat{\psi}^\dagger_i \hat{\psi}_i \hat{\psi}^\dagger_j
  \hat{\psi}_j  $, are decoupled in the standard mean-field way as
  (written schematically)
\begin{gather*}
\rightarrow \hat{\psi}^\dagger_i \hat{\psi}_i \order{ \hat{\psi}^\dagger_j
  \hat{\psi}_j } + \order{ \hat{\psi}^\dagger_i
  \hat{\psi}_i }\hat{\psi}^\dagger _j  \hat{\psi}_j  - \order{ \hat{\psi}^\dagger_i
  \hat{\psi}_i } \order{ \hat{\psi}^\dagger_j
  \hat{\psi}_j }, \\
\rightarrow -\hat{\psi}^\dagger_i \hat{\psi}_j \order{ \hat{\psi}^\dagger_j
  \hat{\psi}_i } - \order{ \hat{\psi}^\dagger_i
  \hat{\psi}_j }\hat{\psi}^\dagger_j \hat{\psi}_i + \order{ \hat{\psi}^\dagger_i
  \hat{\psi}_j } \order{ \hat{\psi}^\dagger_j
  \hat{\psi}_j },
\end{gather*}
the first line representing charge density order and the second bond
density order. 

In terms of the actual superlattice electrons, the charge density
order parameter is defined as
\begin{gather}
\rho_{\alpha i}(l) = \frac{1}{N}\sum_{\vec{k}}\order{\crea{\alpha i}{l,\vec{k}} \anni{\alpha i}{l,\vec{k}} }.
\end{gather}
Bond order parameters are defined as straightforward generalizations
of Ref.~\onlinecite{grushin13}. Of particular interest in our case is
the QH order parameter, constructed from NNN bond order and
defined by Eqs.~\eqref{eq:nnnbond} and~\eqref{eq:qahop}. To obtain
Eq.~\eqref{eq:nnnbond} we decompose
$\mathcal{Y}^{\alpha}_{il,jl'}(\vet{k}-\vec{k}')$, wich arises due to
the reordering of~\eqref{eq:hamv2}, as
\begin{gather}
\mathcal{Y}^{\alpha}_{il,jl'}(\vet{k}-\vec{k}')  = \sum_{\mu =1 }^3 \mathcal{Q}^{\alpha\mu}_{il,jl'}(\vet{k}) \mathcal{Q}^{\alpha\mu}_{il,jl'}(-\vec{k}').
\end{gather}
The sum over $\mu$ follows from the three ways in which NNN sites may
be connected. Note that $\mathcal{Q}^{\alpha\mu}_{il,jl'}(-\vec{k})  =
\mathcal{Q}^{\alpha\mu *}_{il,jl'}(\vec{k})$. We can now NNN bond
order parameters used to construct the QH order parameter of
Eq.~\eqref{eq:qahop}. The NNN bond order mean-fields are defined by
\begin{gather}
\chi^\alpha_{il,jl'}  =  \frac{1}{N}\sum_{\vec{k}} \mathcal{Q}^{\alpha
  *}_{il,jl'} (\vec{k})\order{\crea{\alpha i}{l,\vec{k}} \anni{\alpha j}{l',\vec{k}} }.
\end{gather}
The QH order paramater $\Delta_{\text{QH}}(l)$ is defined so that
each unit cell labeled by $l$ is associated with $2 \times 9$ NNN
bonds. 


\end{document}